\documentclass[final,3p,times,authoryear]{elsarticle}

\usepackage[hidelinks]{hyperref}
\usepackage{amssymb}

\usepackage{amsmath,amsfonts}
\usepackage{algorithmic}
\usepackage{array}
\usepackage{textcomp}
\usepackage{stfloats}
\usepackage{url}
\usepackage{makecell}
\usepackage{verbatim}
\usepackage{graphicx}
\usepackage{multirow}
\usepackage{subfigure}  
\usepackage{makecell}
\usepackage{longtable}
\usepackage{hyperref}
\usepackage{subfigure}  
\usepackage{subcaption}
\usepackage{subfigure}  
\usepackage[ruled,linesnumbered]{algorithm2e}
\usepackage{makecell}
\usepackage{mathrsfs}
\usepackage{xcolor}
\usepackage{enumitem}
\usepackage{rotating}
\usepackage[linesnumbered,ruled]{algorithm2e}

\begin{document}
\begin{frontmatter}
 \title{The Swarm Intelligence Freeway-Urban Trajectories (SWIFTraj) Dataset - Part I: Dataset Description and Applications}

\author[inst1]{Yu Han}
\ead{yuhan2@hkust-gz.edu.cn}

\author[inst2]{Xinkai Ji}

\author[inst2]{Chen Qian}

\author[inst3]{Le Zhang}

\author[inst4,inst5]{Ying Yang}

\author[inst2]{Pan Liu}


\affiliation[inst1]{organization={Thrust of Intelligent Transportation},
            addressline={The Hong Kong University of Science and Technology (Guangzhou)}, 
            city={Guangzhou},
            country={China}}

\affiliation[inst2]{organization={School of Transportation},
            addressline={Southeast University}, 
            city={Nanjing},
            country={China}}

\affiliation[inst3]{organization={School of Economics and Management},
            addressline={Nanjing University of Science and Technology}, 
            city={Nanjing},
            country={China}}

\affiliation[inst4]{organization={School of Management},
            addressline={Shanghai University}, 
            city={Shanghai},
            country={China}}

\affiliation[inst5]{organization={Department of Architecture and Civil Engineering},
            addressline={Chalmers University of Technology}, 
            city={Gothenburg},
            country={Sweden}}
            


\begin{abstract}

This paper presents a detailed description and characterization of a new open-source vehicle trajectory dataset, namely SWIFTraj, constructed from videos recorded by a swarm of 16 drones equipped with 5.4K-resolution cameras. The dataset is distinguished from existing open-source trajectory datasets in several aspects. First, it provides long-distance continuous trajectories of up to 4.5 km on a freeway, enabling in-depth investigation of traffic phenomena and their spatial and temporal evolution. Second, the data collection site covers an integrated network consisting of a long freeway corridor and parts of its connected urban network, facilitating traffic analysis and modeling from a network perspective. The potential applications of the dataset for transportation research, including traffic flow analysis, modeling, and control at multiple scales, as well as topics related to autonomous driving, are thoroughly discussed. Finally, SWIFTraj is released as a freely accessible open-source dataset to support and accelerate future research in the transportation community. The dataset is publicly available at the SWIFTraj website (\href{https://www.swiftraj.com}{https://www.swiftraj.com}).

\end{abstract}


\end{frontmatter}

\section{Introduction}

Trajectory data play a crucial role in advancing traffic flow research. By providing detailed information on individual vehicle movements, they enable the analysis of driving behaviors, the study of complex vehicle interactions, and the linkage between microscopic and macroscopic traffic dynamics. These insights are essential for understanding the mechanisms behind traffic phenomena, developing advanced simulation models, and designing effective traffic control strategies.

Conventional traffic studies have largely relied on aggregated data such as counts and speeds measured by fixed-location sensors \citep{treiber2000congested, cassidy1999some, kerner1996experimental}. While such data have revealed important traffic flow phenomena, such as traffic breakdown, shockwave propagation, and capacity drop, they cannot capture the underlying mechanisms at the vehicle level. Moreover, aggregated measurements fall short in addressing the uncertainties of traffic flow dynamics. Predicting the exact timing of breakdowns or the extent of capacity drop remains difficult without considering heterogeneous driving behaviors and complex vehicle interactions. These limitations highlight the need for high-resolution trajectory data that can provide deeper insights into traffic complexities.

The release of the NGSIM dataset nearly two decades ago marked a milestone in traffic flow research. It has spurred extensive studies on driver behavior, traffic phenomena, and the development of traffic flow models at different scales \citep{li2020trajectory}. Nevertheless, the NGSIM dataset has several limitations. Its data quality has been criticized, with issues such as unrealistically large accelerations and apparent trajectory collisions after accounting for vehicle length \citep{coifman2017critical}. In addition, its spatial and temporal coverage is restricted. Temporally, it does not capture the transition from free flow to congestion; spatially, it only covers a limited portion of the downstream of congestion. These shortcomings make the dataset insufficient for in-depth analysis of certain traffic phenomena, such as breakdown and capacity drop.

In recent years, high-resolution trajectory datasets have become increasingly available, extending the applicability of trajectory data beyond what NGSIM offered. For example, the Zen Traffic dataset provides broader spatial and temporal coverage of freeway traffic \citep{seo2020evaluation}, enabling more in-depth analysis of active bottlenecks. The pNEUMA dataset focuses on urban networks with multi-modal traffic flows \citep{barmpounakis2020new}, supporting network-level studies of urban traffic. The CitySim dataset offers trajectories that capture a substantial number of critical safety events across diverse scenarios, including intersections as well as freeway merge, diverge, and weaving sections, thereby facilitating safety-oriented research \citep{zheng2024citysim}.

This paper introduces a new open-source trajectory dataset, namely SWIFTraj (Swarm Intelligence Freeway–Urban Trajectories). The data were collected on a road network in Nanjing, China, using a swarm of 16 drones for recording and object detection algorithms for extracting vehicle trajectories. The largest portion of the dataset was collected from an integrated network consisting of a 4.5 km freeway stretch and its connected urban network (about 2 km²). The SWIFTraj dataset is distinguished by the following aspects:
(i) Continuous, long-distance trajectories. Each vehicle is assigned a unique identifier, enabling its full trajectory to be tracked from entry to exit within the study area. This provides sufficient spatial and temporal coverage of typical freeway active bottlenecks, making the dataset a valuable complement to existing state-of-the-art trajectory datasets.
(ii) Integrated freeway–urban network. The study area includes a 4.5 km freeway stretch and its connected urban roads, supporting applications in both freeway traffic analysis and network-level modeling.

The main contribution of this article is the creation of the SWIFTraj dataset, released as an open resource for the research community. We describe its key components, including the study area and road network geometry, the dataset features, and its potential applications. In a companion paper, we introduce the technologies used to connect trajectories across consecutive images, enabling the construction of long-distance, continuous vehicle trajectories.

The remainder of this paper is organized as follows. Section 2 reviews existing trajectory datasets and their applicability. Section 3 describes the experimental design. Section 4 discusses potential applications of the dataset. Section 5 concludes the paper.

\section{Related works}
\label{sec:related_work}

Traffic data collection has been extensively studied in the literature. Since this paper focuses on trajectory data, this section exclusively reviews existing open-source trajectory datasets that provide complete vehicle trajectories without sampling. Conventional data sources for macroscopic traffic states (e.g., flow, speed, and occupancy) and sampled trajectory data from probe vehicles equipped with on-board detectors or GNSS devices are therefore beyond the scope of this review.

There are more than 20 open-source trajectory datasets reported in the literature, most of which have been released within the past five years. Table 1 summarized existing open-source trajectory datasets that we have found. The NGSIM dataset \citep{alexiadis2004next}, which was released about two decades ago, is among the earliest open-source trajectory datasets that provide complete vehicle trajectories. The NGSIM dataset has several limitations. Its data quality has been criticized, with issues such as unrealistically large accelerations and apparent trajectory collisions after accounting for vehicle length \citep{coifman2017critical}. In addition, its spatial and temporal coverage is restricted. Temporally, it does not capture the transition from free flow to congestion; spatially, it only covers a limited portion of the downstream of congestion. These shortcomings make the dataset insufficient for in-depth analysis of certain traffic phenomena, such as breakdown and capacity drop.

The leveL-X Data project, developed by a German technology company, comprises a series of datasets covering different road environments, including homogeneous freeway segments (HighD) \citep{krajewski2018highd}, merge and diverge sections (ExitD) \citep{exiDdataset}, urban intersections (InD), roundabouts (RounD), and mixed traffic flows (UniD). All trajectories in these datasets were captured using a single UAV camera, which limits their spatial coverage to approximately 400–500 meters. The leveL-X datasets support not only the analysis of traffic behavior but also the testing and validation of autonomous driving systems \citep{geng2023multimodal}.

The Zen Traffic Dataset contains continuous vehicle trajectories collected from the Hanshin Expressway, a two-lane urban expressway in Osaka, Japan \citep{seo2020evaluation}. The data cover both merge and sag bottlenecks and were captured using 38 roadside cameras. Due to the limited fields of view of roadside cameras, some vehicles were occluded and could not be directly observed; trajectories of these vehicles were reconstructed through interpolation. The Zen Traffic Dataset has supported advanced research in traffic state estimation, traffic flow modeling, and dynamic traffic control \citep{he2025constructing, he2025review, liu2025optimizing}.

The pNEUMA dataset is the first large-scale complete vehicle trajectory dataset for an urban road network. The data were collected from a 1.3 km² congested downtown area in Athens using a swarm of 10 UAVs. The dataset covers approximately 100 busy intersections and more than 30 bus stops. The trajectories are continuous and include multimodal traffic. Compared with earlier datasets that primarily focus on freeway segments or small-scale sites, pNEUMA substantially extends the spatial coverage and captures the complexity of dense urban traffic environments. It has facilitated research on vehicle interaction mechanisms, lane choice behavior on urban arterial, and the development of traffic control strategies such as eco-driving \citep{jiao2023inferring, li2025trajpt, yang2024eco}.

The CitySim dataset provides vehicle trajectories collected from multiple sites in China and the United States. It is designed to capture critical safety events in conflict zones, including various types of urban intersections as well as freeway merge, diverge, and weaving sections, thereby facilitating safety-oriented research \citep{zheng2024citysim}. The dataset also supports digital-twin-related studies by providing three-dimensional base maps of the recording locations along with traffic signal timing information. Most scenarios were collected using single UAVs with limited spatial coverage; only a few sites employed two UAVs, extending the observation range to approximately 1 km. 

The I24-Motion dataset contains large-scale vehicle trajectories collected from a 6.75 km stretch of the I-24 freeway near Nashville, USA, using 276 roadside cameras \citep{gloudemans202324}. While it is the largest open-source trajectory dataset available to date, its main limitation is that the trajectories are not continuous. This discontinuity restricts its applicability for analyzing the variation of driving behavior in complex traffic contexts, as well as for linking individual driving behaviors with macroscopic traffic phenomena.

As summarized in Table~\ref{tab:comparison_datasets_simple}, existing open-source trajectory datasets exhibit substantial differences in spatial coverage, traffic context, and trajectory continuity. While some datasets provide high-quality complete trajectories at relatively small spatial scales, and others cover broader areas with limited continuity, SWIFTraj distinguishes itself by offering long-range continuous trajectories of up to 4.5~km within an integrated freeway–urban network. These features enable the analysis of vehicle behavior and traffic evolution over extended spatial scales and across more diverse traffic environments, which are difficult to achieve using existing datasets.

\begin{sidewaystable}
\caption{Comparison of Traffic Datasets}
\label{tab:comparison_datasets_simple}
\centering
\begin{tabular}{>{\raggedright\arraybackslash}m{6.0cm} 
                >{\centering\arraybackslash}m{1.0 cm}
                >{\raggedright\arraybackslash}m{3.5cm}
                >{\raggedright\arraybackslash}m{2.5cm}
                >{\raggedright\arraybackslash}m{3.0cm}
                >{\raggedright\arraybackslash}m{2cm}
                >{\centering\arraybackslash}m{1.2cm}}
\hline
\textbf{Dataset} & \textbf{Year} & \textbf{Location} & \textbf{Context} & \textbf{Sensor} & \parbox{1.5cm}{\centering \textbf{Maximum} \\ \textbf{Coverage}} & \parbox{1.5cm}{\centering \textbf{spatial} \\ \textbf{continuity}} \\ 
\hline

\textbf{NGSIM US-101} \\ \citep{alexiadis2004next} & 2005 & Los Angeles, USA & Freeway & Roadside camera & 0.64 $km$ & Yes \\

\textbf{HighD} \\ \citep{krajewski2018highd} & 2018 & Cologne, Germany & Freeway & UAV camera & 0.42 $km$ & Yes \\

\textbf{Zen Traffic Dataset} \\ \citep{seo2020evaluation} & 2018 & Osaka, Japan & Freeway & Roadside camera & 2 $km$ & Yes \\

\textbf{Automatum} \\ \citep{spannaus2021automatum} & 2021 & Germany & Freeway & UAV camera & 0.66 $km$ & Yes \\

\textbf{HIGH-SIM} \\ \citep{shi2021video} & 2021 & Florida, USA & Freeway & Helicopter camera & 2.44 $km$ & Yes \\

\textbf{ExitD} \\ \citep{exiDdataset} & 2021 & Germany & Freeway & UAV camera & 0.42 $km$ & Yes \\

\textbf{pNEUMA} \\ \citep{barmpounakis2020new} & 2020 & Athens, Greece & Urban & UAV camera & 1.3 $km^2$ & Yes \\

\textbf{CitySim} \\ \citep{zheng2024citysim} & 2020 & China and USA & Freeway, Urban & UAV camera & 0.73$km$ & Yes \\

\textbf{MAGIC} \\ \citep{ma2022magic} & 2022 & Shanghai, China & Freeway & UAV camera & 2 $km$ & No \\

\textbf{I24-motion} \\ \citep{gloudemans202324} & 2022 & Nashville, USA & Freeway & Roadside camera & 6.75 $km$ & No \\

\textbf{Ubiquitous Traffic Eyes} \\ \citep{feng2025ubiquitous} & 2025 & Nanjing, China & Freeway & UAV camera & 0.38 $km$ & Yes \\

\textbf{TOD-VT} \\ \citep{wen2024safety} & 2024 & Wuhan, China & Freeway & UAV camera & 1.2 $km$  & Yes \\

\textbf{INTERACTION} \\ \citep{zhan2019interaction} & 2019 & China, Germany, USA & Freeway, Urban & UAV camera & 0.12 $km$  & Yes \\

\textbf{Songdo Traffic} \\ \citep{fonod2025advanced} & 2025 & South Korea & Urban & UAV camera &  0.78 $km^2$  & No \\

\textbf{CQSkyEyeX} \\ \citep{xu2022cqskyeyex} & 2022 & China & Freeway & UAV camera &   0.4 $km$  & Yes \\

\textbf{AD4CHE} \\ \citep{zhang2023ad4che} & 2023 & China & Freeway & UAV camera &    0.14$km$  & Yes \\

\textbf{ISAC} \\ \citep{berghaus2024vehicle} & 2024 & Germany & Freeway & UAV camera &  1.13 $km$  & Yes \\

\textbf{MiTra} \\ \citep{chaudhari2025mitra} & 2025 & Germany & Freeway & UAV camera &   0.88 $km$  & Yes \\

\textbf{HDSVT} \\ \citep{wen2025hdsvt} & 2025 & China & Bridge & UAV camera & 7 $\times$ 0.20 $km$  & No \\

\textbf{SWIFTraj} & 2025 & Nanjing, China & Freeway-Urban integration & UAV camera & 4.5 km & Yes \\ \hline
\end{tabular}
\end{sidewaystable}

\section{Experiment description}
\label{sec:experiment}

\subsection{Experiment design}

\begin{figure}[!ht]
	\centering
 	\includegraphics[width=0.9\textwidth]{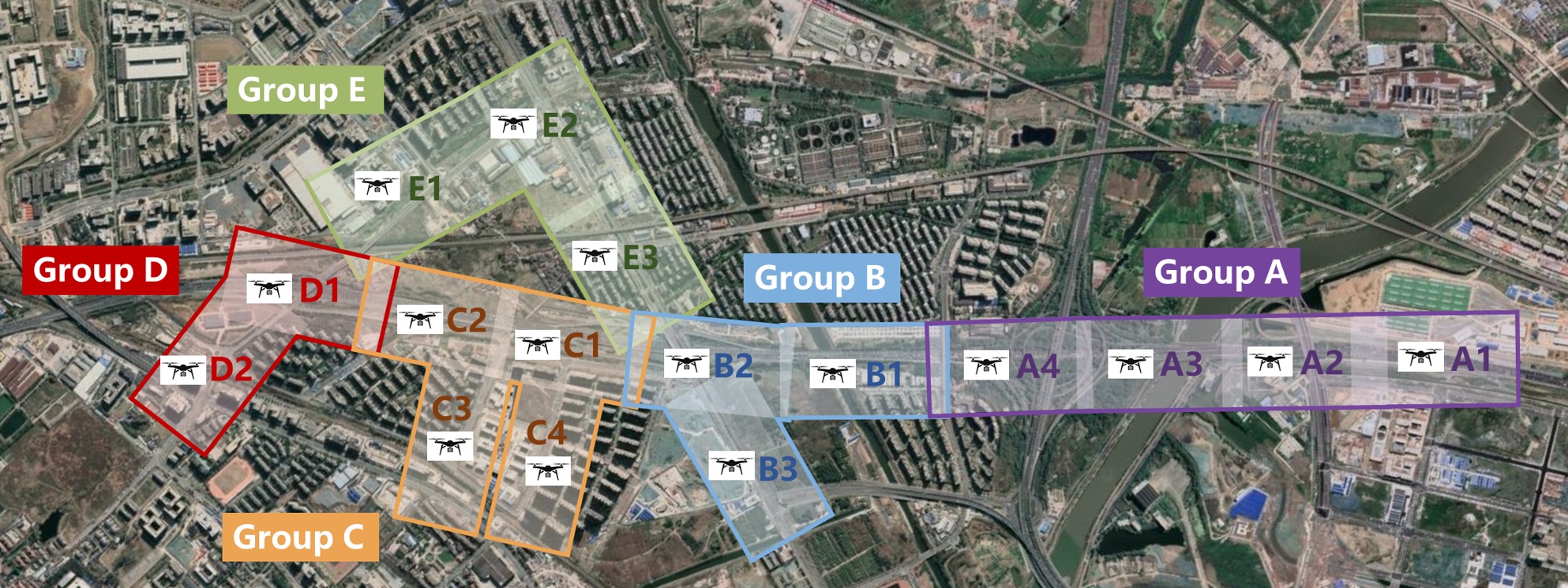}
	\caption{Satellite map of the site for the data collection experiment.}
	\label{fig:map}
\end{figure}

The data collection experiment was conducted on a road network in Nanjing, China, during the morning peak hours on two sunny workdays, June 16 and 17, 2022, following a trial run on June 15. The study area, shown in the satellite map in Fig. \ref{fig:map}, was covered using 16 drones, each monitoring a region of approximately 600 m × 340 m. Overlapping fields of view between neighboring drones ensured that trajectories from neighboring UAV views could be connected into continuous vehicle trajectories. The dataset covers multiple traffic states during the morning peak period, including pre-peak free-flow conditions, the transition from free flow to congestion, and the dissipation of congestion. In addition, owing to the large spatial coverage of the dataset, heterogeneous traffic states can coexist simultaneously across different road segments, with some segments operating under free-flow conditions while others experience congestion.

Sixteen experienced pilots, divided into five groups (A–E), operated the drones. Each group was led by a captain who coordinated battery changes both within and across groups to maximize overlap in recording time. Each pilot was equipped with five batteries, with each battery sustaining a drone for approximately 20 minutes. Depending on the distance between the pilot and the drone’s hovering point, the effective recording duration of each flight ranged from 12 to 18 minutes, and battery replacement required about 5 minutes. Each day’s experiment began at 7:00 am and ended around 9:00 am.

During the trial run, the operating location of each pilot and the corresponding hovering point of each drone were determined. The operating locations were chosen at spacious sites to avoid obstruction during flight and to ensure pilot safety. In addition, the flying distances between each pilot and the hovering point of the corresponding drone were kept as small as possible to facilitate better synchronization of recording time.

\begin{figure}[!ht]
	\centering
 	\includegraphics[width=0.9\textwidth]{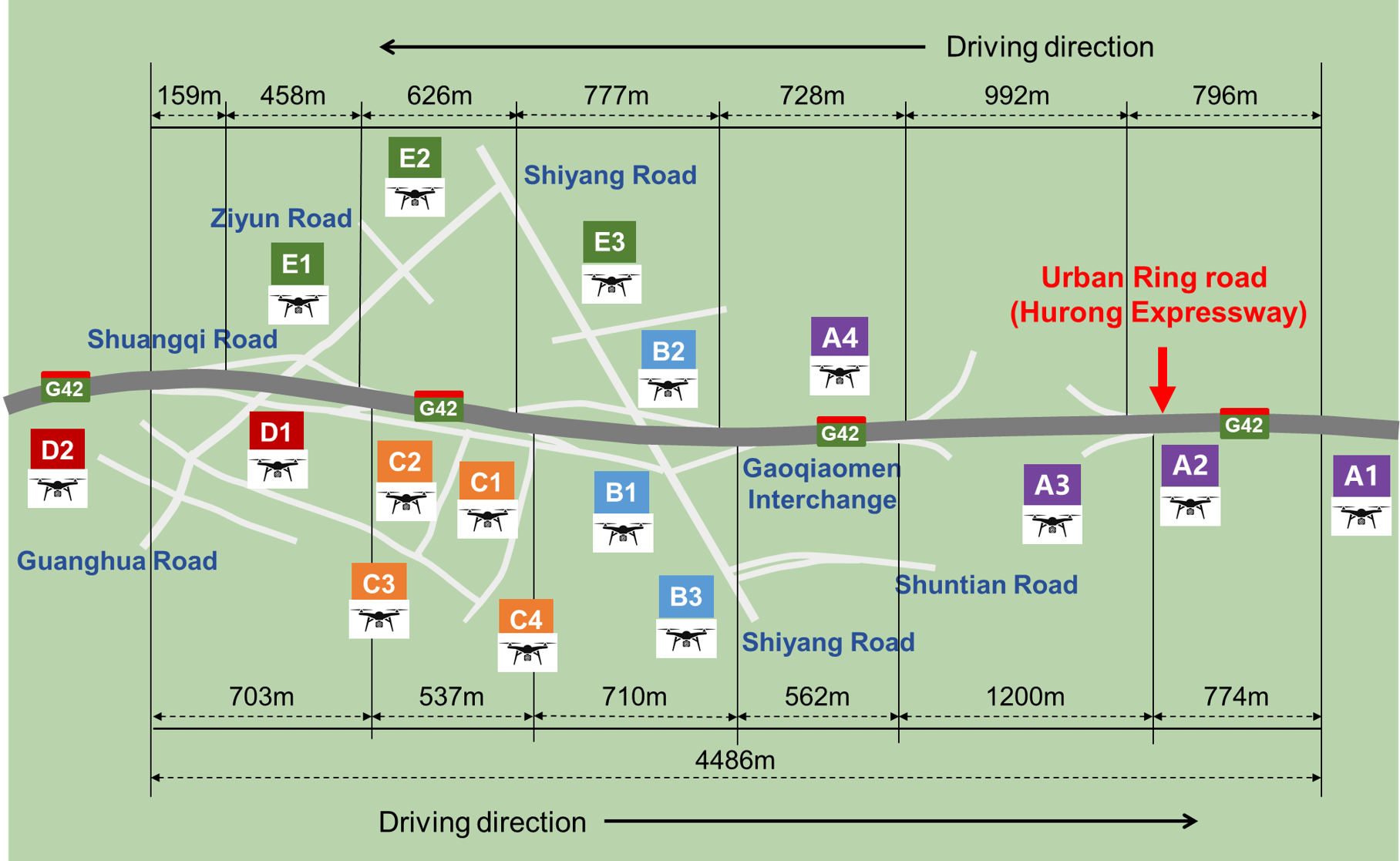}
	\caption{Topological map of the site for the data collection experiment.}
	\label{fig:topology}
\end{figure}

The site covers a 4.5 km stretch of the urban ring road (the Hurong Expressway) and part of its connected urban network. As shown in the topological map in \ref{fig:topology}, traffic on the freeway is two-directional. In the left-to-right direction, two active bottlenecks typically occur during the morning peak hours. One is located at a merge section where traffic enters from an urban arterial, Shiyang Road, and the other occurs at the merge with another freeway at the Gaoqiaomen Interchange. In the opposite (right-to-left) direction, a single active bottleneck is usually observed, originating from the merge section at the Gaoqiaomen Interchange.

\subsection{Trajectory processing}

All trajectory data were collected using DJI AIR 2S drones equipped with high-resolution 5.4K cameras (5472 × 3078 pixels). The recorded videos served as the raw input for subsequent trajectory processing. Expressway traffic trajectories were extracted from videos captured by nine UAV deployments (A1–A4, B1–B2, C1–C2, and D1). Trajectory processing generally involves three main stages: trajectory extraction, trajectory reconstruction under occlusion, and trajectory stitching across consecutive images.

\begin{figure}[!ht]
	\centering
 	\includegraphics[width=0.8\textwidth]{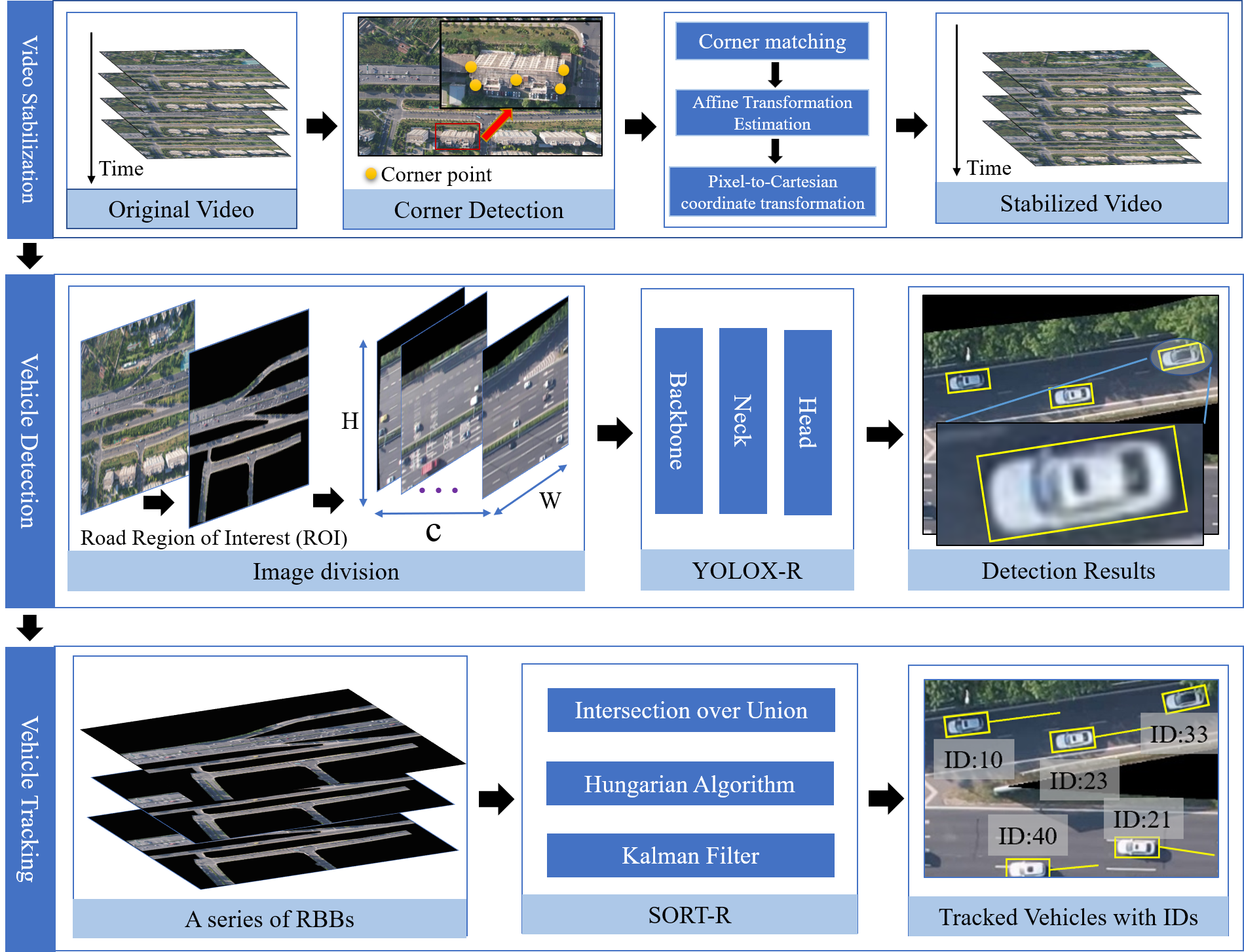}
	\caption{Overview of the UAV-based vehicle trajectory extraction pipeline.}
	\label{fig:extraction}
\end{figure}

For trajectory extraction, we adopt OpenVTER \citep{ji2024openvter}, a generalized open-source vehicle trajectory extraction framework based on rotated bounding boxes (RBBs). The overall processing pipeline is illustrated in Fig.~\ref{fig:extraction} and consists of three main steps: video stabilization, vehicle detection, and vehicle tracking. In the first step, a base-frame video stabilization strategy is applied to mitigate cumulative errors in the homography transformation and improve computational efficiency. Next, to improve detection of small, densely packed vehicles in high-resolution UAV imagery, each stabilized frame is cropped to the road region of interest and further divided into multiple sub-images. Vehicle detection is then performed on these sub-images using YOLOX-R, a rotated object detection model that can accurately identify vehicles with arbitrary orientations through RBBs. The detected RBBs are subsequently mapped back to the original image coordinate system. Finally, a rotated vehicle tracking algorithm, SORT-R, is used to associate vehicle detections across consecutive frames. SORT-R integrates intersection-over-union (IoU) matching, the Hungarian assignment algorithm, and Kalman filtering to achieve real-time multi-vehicle tracking. Vehicle trajectories are then constructed as time-ordered sequences of tracked RBB centers.

Due to the presence of overpass bridges within the data collection area, the mainline roadway experiences partial occlusions, leading to intermittent losses in vehicle tracking and fragmented trajectories. To ensure trajectory continuity along the roadway, the missing trajectory segments within these bridge-occluded regions must be reconstructed. Trajectory reconstruction is formulated as a two-step problem: (1) matching vehicles before and after the occlusion, and (2) generating trajectories within the occluded region for the matched vehicles. We adopt a hybrid macro–micro framework to reconstruct the missing trajectories, such that the reconstructed trajectories are consistent with both macroscopic speed estimates derived from traffic-wave propagation and microscopic kinematic constraints of individual vehicle motion.

\begin{figure}[!ht]
	\centering
 	\includegraphics[width=0.8\textwidth]{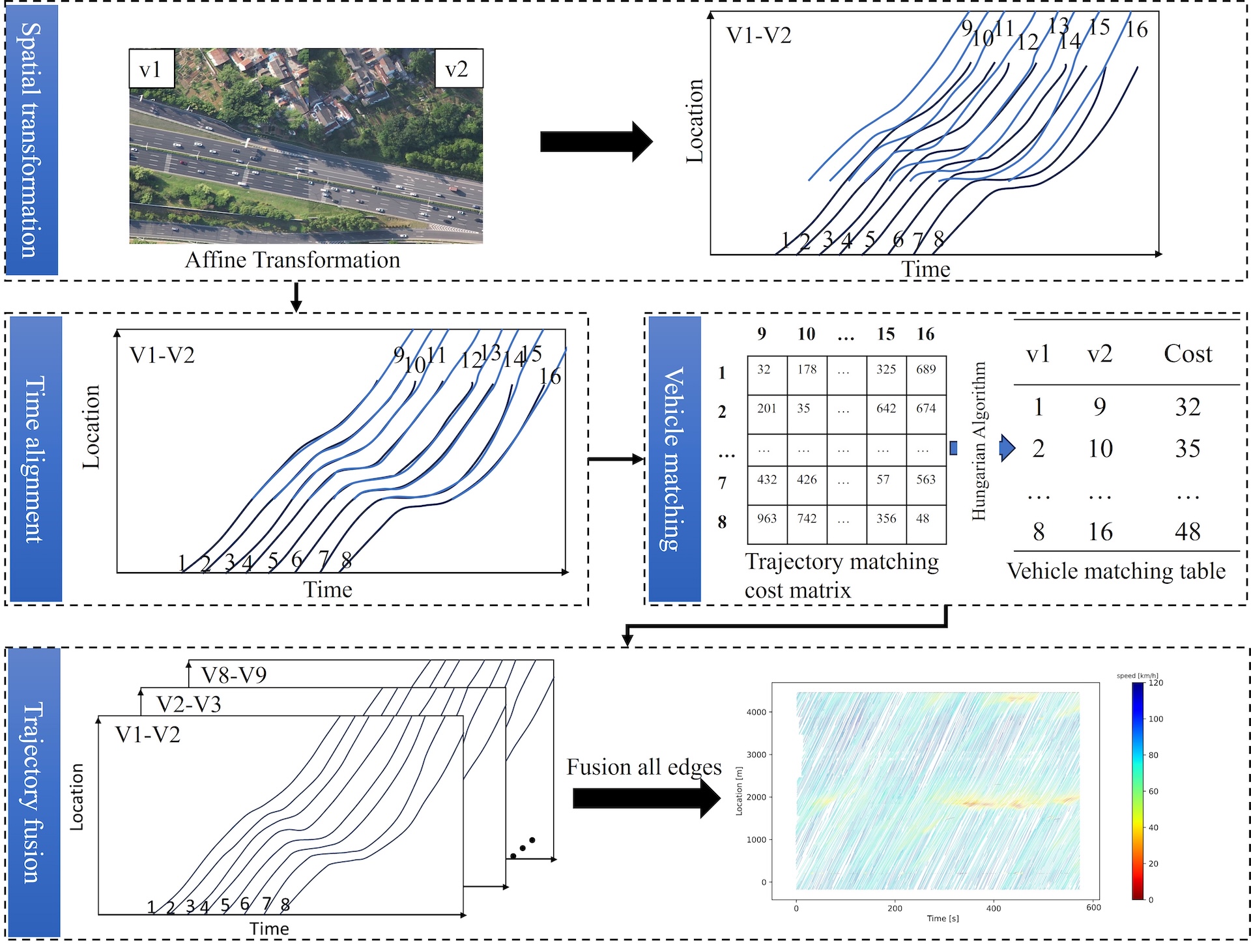}\label{fig:traj_process:b}
	\caption{Overview of vehicle trajectory connection pipeline.}
	\label{fig:stitch}
\end{figure}

After trajectory reconstruction, continuous vehicle trajectories are obtained within each individual video. To further achieve continuity across multiple videos, trajectory stitching is performed to associate trajectories observed by adjacent UAVs within overlapping regions. The objective is to merge trajectory fragments belonging to the same vehicle across different UAV views, thereby resolving discontinuities caused by limited camera coverage and ensuring spatiotemporal consistency. The overall trajectory stitching pipeline, illustrated in Figure~\ref{fig:stitch}, consists of four main steps: spatial transformation, time alignment, vehicle matching, and trajectory fusion. Specifically, a graph is constructed to represent the UAV layout, where each video is treated as a node and edges are defined based on overlapping neighboring views. This graph structure enables systematic trajectory stitching under irregular UAV spatial distributions. Trajectories from different UAV coordinate systems are first transformed into a unified spatial reference frame. Temporal offsets between adjacent videos are then estimated through trajectory matching cost minimization and corrected to ensure temporal consistency. Based on the aligned trajectories in overlapping regions, cross-video vehicle association is performed using the Hungarian algorithm. Finally, matched trajectory fragments are fused to generate continuous trajectories across multiple UAV views while maintaining temporal vehicle ID consistency. More detailed algorithmic derivations and implementation details are provided in the companion paper.

\subsection{Trajectory evaluation}

The quality of the trajectory data was evaluated using two performance metrics, namely the complete trajectory ratio (CTR) and internal consistency. Specifically, we computed the CTR for videos collected at mainline expressway sites. The CTR is used to evaluate the completeness of extracted vehicle trajectories and is defined as:

\begin{equation}
\mathrm{CTR} = \frac{n_c}{n_t},
\end{equation}
where $n_c$ is the number of complete vehicle trajectories and $n_t$ is the total number of mainline vehicle trajectories. A complete trajectory is defined as one in which a vehicle is continuously observed throughout the study road segment without false interruption. A higher CTR indicates better performance in preserving trajectory continuity and maintaining consistent vehicle identities over time.

In addition, we conducted a quantitative evaluation of trajectory quality based on an internal consistency analysis of the extracted position and speed data. Specifically, we assess whether derived trajectory variables remain mutually consistent over time. The internal consistency of position is defined as:

\begin{equation}
\epsilon_t^{s} = \hat{s}(t) - \left( \hat{s}(0) + \int_{0}^{t} \hat{v}(\tau)\, d\tau \right),
\end{equation}

\begin{equation}
\epsilon_t^{v} = \hat{v}(t) - \left( \hat{v}(0) + \int_{0}^{t} \hat{a}(\tau)\, d\tau \right),
\end{equation}

\noindent
where $\epsilon_t^{s}$ and $\epsilon_t^{v}$ denote the internal consistency errors of position and speed at time $t$, respectively. 
$\hat{s}(t)$ and $\hat{s}(0)$ represent the observed positions of the vehicle at time $t$ and the initial time. 
$\hat{v}(t)$ and $\hat{v}(0)$ denote the observed instantaneous speeds at time $t$ and the initial time, respectively. 
$\hat{v}(\tau)$ and $\hat{a}(\tau)$ are the observed instantaneous speed and acceleration at time $\tau$. 
The integrals $\int_{0}^{t} \hat{v}(\tau)\, d\tau$ and $\int_{0}^{t} \hat{a}(\tau)\, d\tau$ represent the position and speed reconstructed from the observed speed and acceleration over time.

We analyzed the CTR for each individual site. Across the nine mainline expressway video sites, covering eight lanes in both directions, a total of 8,242 vehicle trajectories from a single flight were used for quality assessment. Among these, 8,067 were classified as complete trajectories and 175 as incomplete trajectories, resulting in an average CTR of 97.88\%. This result indicates that the vast majority of vehicle trajectories can be continuously reconstructed without false interruption, demonstrating strong trajectory continuity and temporal vehicle ID consistency.

To further evaluate noise levels and error characteristics, we conducted the internal consistency analysis for both position and speed. Across the nine sites, the mean absolute internal consistency error is 0.031~m for position and 0.016~m/s for speed. The position consistency ranges from $-0.193$~m to $0.601$~m, while the speed consistency ranges from $-0.293$~m/s to $0.223$~m/s. These results indicate that the extracted trajectories are generally smooth and physically consistent, although a small number of trajectories exhibit relatively large local deviations.


\section{Data description}

\begin{figure}[!ht]
	\centering
 	\includegraphics[width=0.8\textwidth]{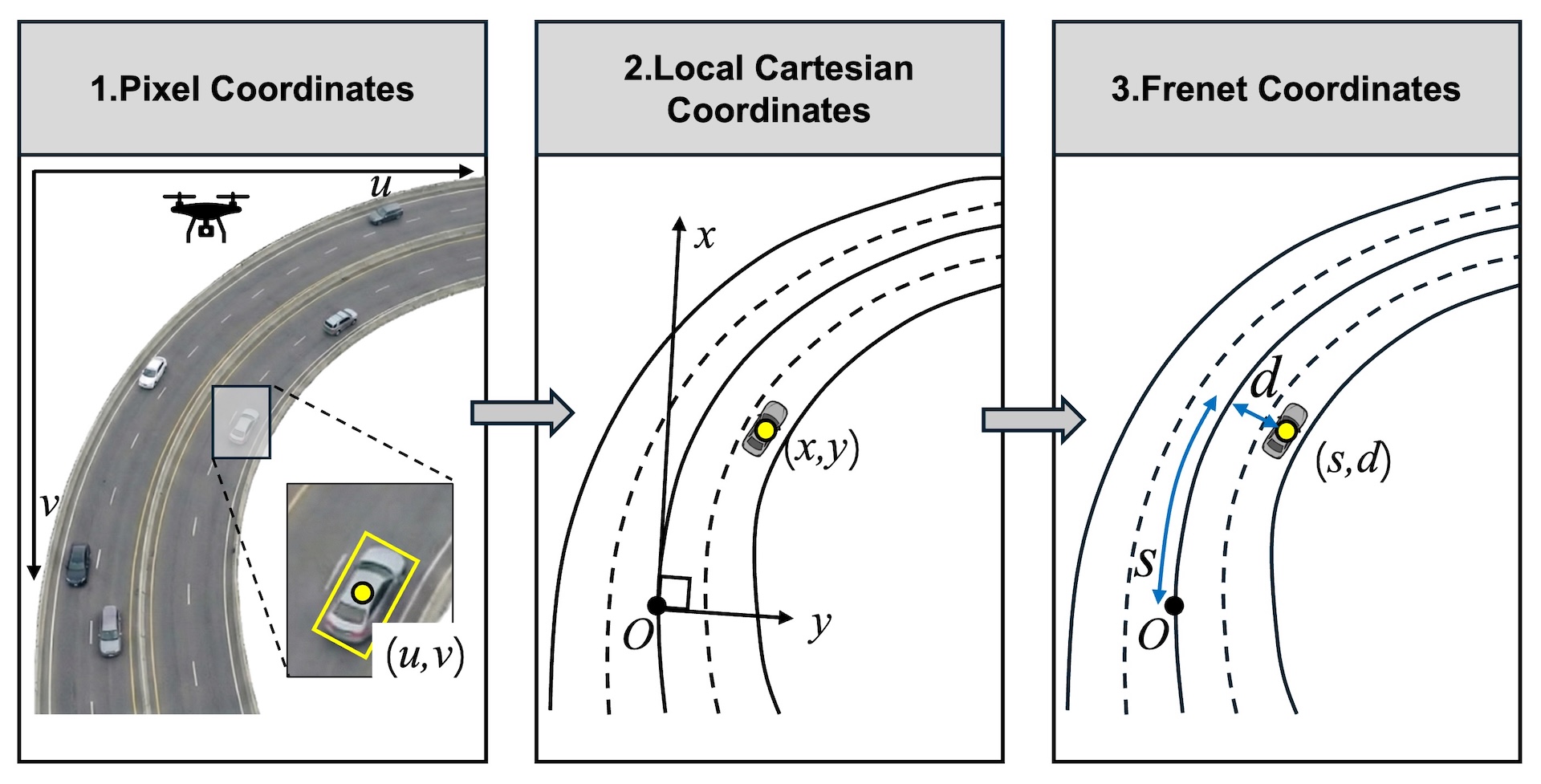}
	\caption{The process of coordination transformation.}
	\label{fig:coordinate}
\end{figure}

Vehicle trajectories are extracted from UAV videos through a multi-stage coordinate transformation process, as illustrated in Figure~\ref{fig:coordinate}. First, vehicle positions are detected and tracked frame by frame using object detection and multi-object tracking algorithms, yielding pixel-level coordinates $(u, v)$ in the image space.

To map pixel coordinates to the physical world, a spatial scale is established using standard dashed lane markings on the roadway, whose actual length is 6 meters. Based on this scale, a local Cartesian coordinate system is defined by selecting a reference point as the origin. The $x$-axis is aligned with the local tangent direction of the roadway, corresponding to the nominal vehicle travel direction, and the $y$-axis is defined as the perpendicular lateral direction. After coordinate translation and axis alignment, pixel coordinates are converted into physical Cartesian coordinates $(x, y)$ in meters.

Although the Cartesian coordinates provide metric vehicle positions, a fixed coordinate orientation is insufficient to represent motion along curved roadways. Therefore, a Frenet coordinate system is further introduced to describe vehicle motion relative to the lane geometry. The lane centerline is extracted as the reference curve, and vehicle positions are projected onto this curve to obtain the longitudinal coordinate $s$, defined as the arc length along the centerline, and the lateral coordinate $d$, defined as the perpendicular offset from the centerline. The coordinate $s$ characterizes longitudinal movement along the roadway, while $d$ captures lateral movement across lanes. The dataset provides vehicle trajectories in both Cartesian $(x, y)$ and Frenet $(s, d)$ coordinate systems, supporting geometric analysis as well as behavior-oriented traffic flow modeling.

\begin{figure}[!ht]
	\centering
	\includegraphics[width=0.95\textwidth]{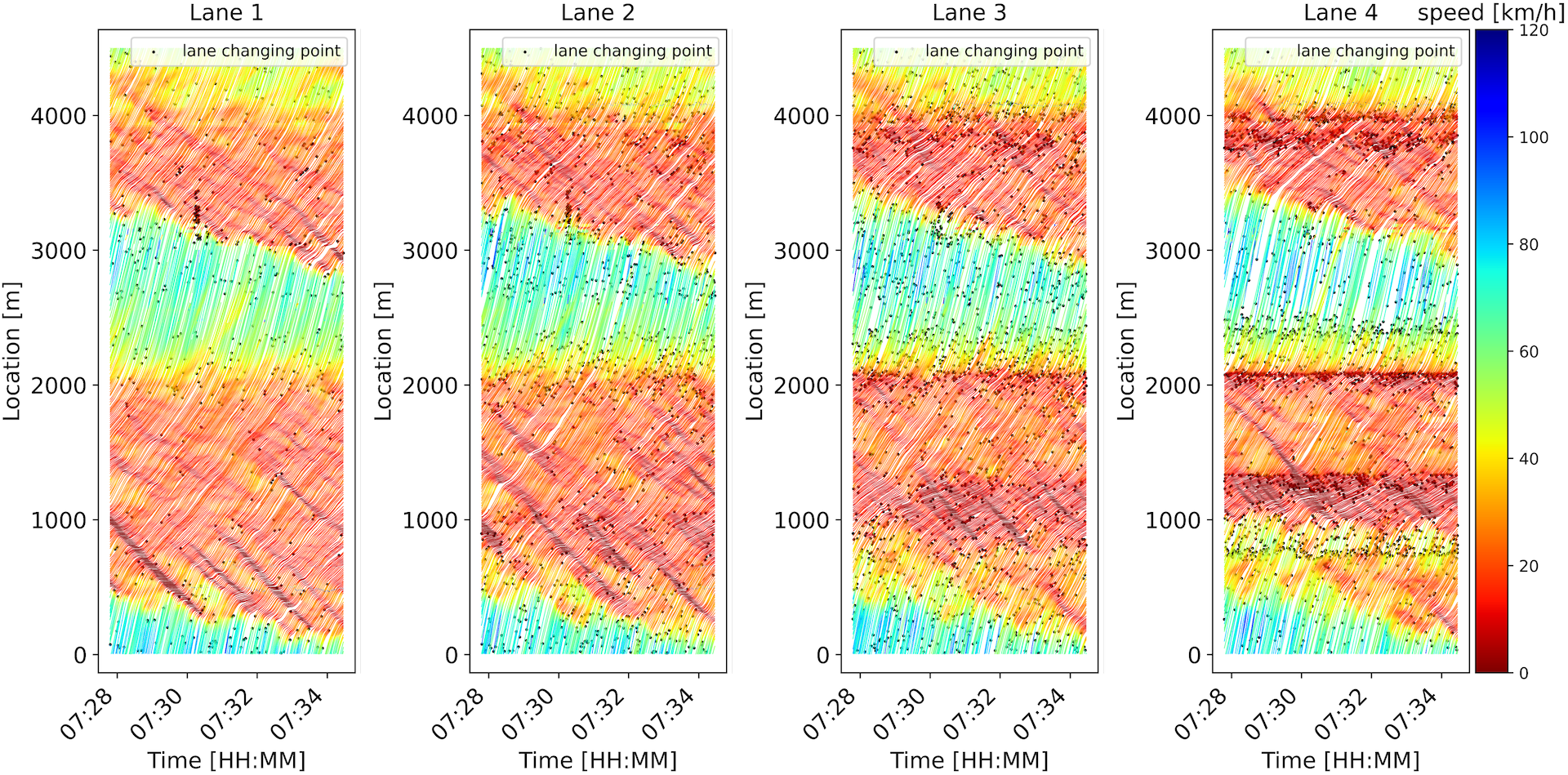}
	\caption{Color-coded trajectories on the freeway for the left-to-right direction. The lane change points are highlighted with black marks.}\label{fig:traj_example}
\end{figure}

For freeway traffic, videos recorded by the nine UAVs operating within the same time slot are processed using the software pipeline described in Section~3. After processing, each vehicle on the freeway, including on-ramp vehicles, is assigned a unique identifier, resulting in continuous vehicle trajectories along the freeway. An example of continuous trajectories for the left-to-right direction is shown in Figure~\ref{fig:traj_example}, where colors indicate vehicle speed.

The data are organized into sub-datasets on a per-video basis. Each video corresponds to a sub-dataset representing vehicle trajectories collected at a specific location during one time slot. For each trajectory, the data fields include vehicle attributes and motion information, such as timestamps and positions. The position data record not only the two-dimensional footprint of the rear-center point of each vehicle, but also the coordinates of the four vehicle corners. Derivative quantities, including velocity, acceleration, and steering angle, can be directly computed from the position data using standard finite-difference methods. Detailed field definitions are provided in Appendix~A.

For vehicle trajectories on the urban network, the data are currently processed only through the trajectory extraction step, without stitching across different images. On one hand, trajectory stitching for urban traffic is more challenging due to the greater diversity of traffic movements compared with freeway traffic. On the other hand, research on the relationship between microscopic vehicle behavior and macroscopic traffic flow characteristics has received relatively less attention in urban contexts than on freeways, where such analyses typically require continuous trajectories with large spatial and temporal coverage. As a result, trajectory stitching for urban traffic remains an ongoing effort. 

\begin{figure*}[!h]
	\centering
	\subfigure[]{\includegraphics[width=0.54\textwidth] {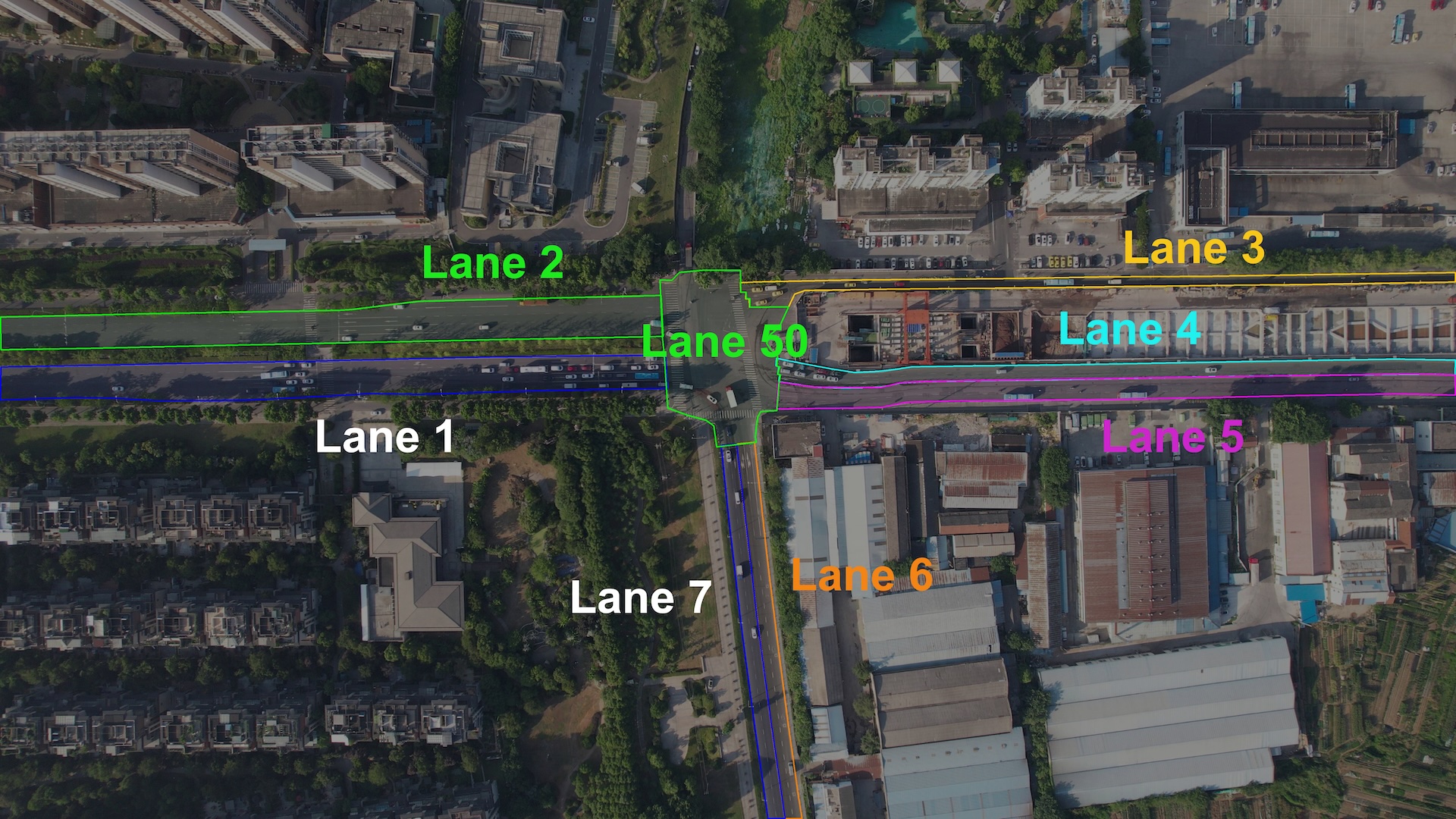}}
	\subfigure[]{\includegraphics[width=0.45\textwidth] {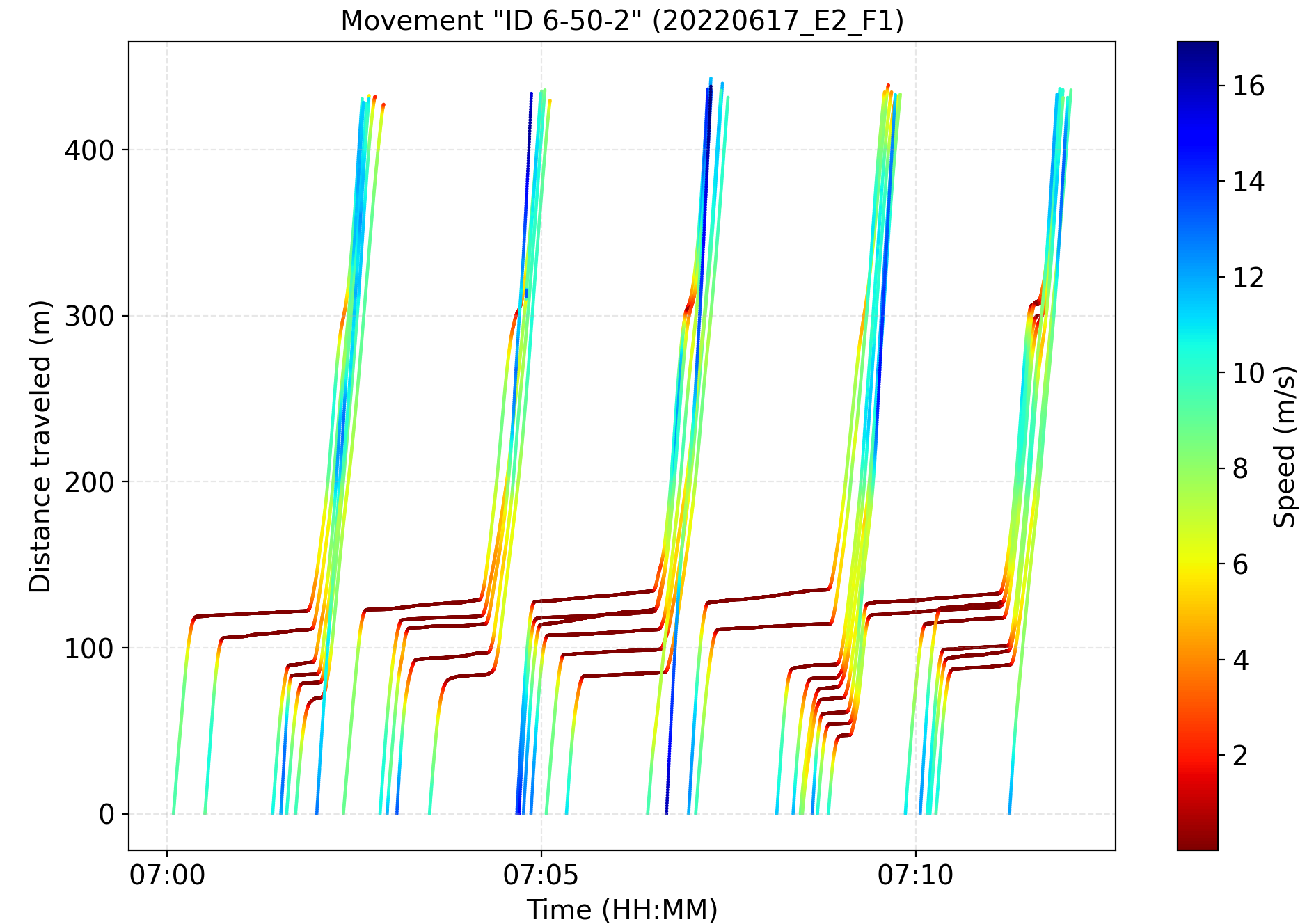}} 
	\caption{An example of an intersection configuration and trajectories corresponding to a left-turn movement.}\label{fig:intersection}
\end{figure*}

The field definitions for urban traffic are largely the same as those for freeway traffic, with one additional field indicating the movement of each trajectory, as detailed in Appendix A. For field name consistency, which simplifies data processing, we use the same field name lane\_{id} to represent either a road section or a node in the urban network. Here, "lane" does not refer to a physical freeway lane, but to an area corresponding to a specific road segment or an intersection. An example of trajectories from an urban intersection at site C3 is presented. The topology of the intersection is shown in Figure~\ref{fig:intersection} (a), where both the intersection (lane 50) and its connected road segments are numbered. Accordingly, a movement ID is defined in the form of “incoming road segment–intersection–outgoing road segment.” In this example, trajectories with movement ID “6–50–2” are illustrated in Figure~\ref{fig:intersection} (b).

\section{Data usage for research possibilities}

This section presents potential applications of the proposed dataset, including traffic phenomena analysis, driving behavior modeling, network traffic flow analysis, and research on connected and autonomous vehicles (CAVs).

\subsection{Analysis of traffic phenomena}

Trajectory data are essential for understanding the mechanisms underlying macroscopic traffic phenomena, such as capacity drop and traffic oscillations \citep{zheng2011freeway, chen2012microscopic, tian2016empirical}. Sufficient spatial and temporal coverage is critical for conducting comprehensive analyses based on trajectory data. For example, investigating capacity drop requires continuous trajectories with temporal coverage spanning the transition from uncongested to congested states, as well as spatial coverage that includes the bottleneck area and its downstream free flow section.

\begin{figure}[!ht]
	\centering
 	\includegraphics[width=0.8\textwidth]{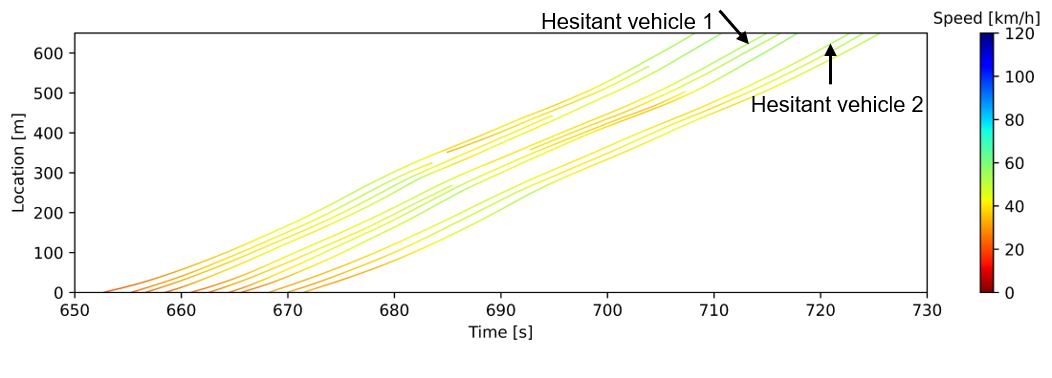}
	\caption{Example of capacity drop analysis using trajectory data, where hesitant vehicles generate voids and cause capacity loss.}
	\label{fig:cdrop}
\end{figure}

Most existing open trajectory datasets listed in Table~\ref{tab:comparison_datasets_simple} do not satisfy these requirements. Some datasets suffer from limited spatial or temporal coverage, such as NGSIM, while others provide discontinuous trajectories, such as the I24-Motion dataset. The Zen Traffic Dataset is among the few datasets that satisfy both spatial and temporal coverage requirements. Nevertheless, the dataset presented in this paper still offers several unique features. The Zen Traffic Dataset focuses on a two-lane freeway with sag sections, whereas our dataset includes typical merge-induced bottlenecks on four-lane freeway segments. Furthermore, driving in Japan follows left-hand traffic, while driving in China follows right-hand traffic. These differences make our dataset attractive to researchers interested in traffic flow analysis. In \citet{han2025capacity}, this dataset was used to support an in-depth analysis of capacity drop. It was found that the late responses of hesitant vehicles during the acceleration process significantly contribute to capacity drop. An illustrative example is shown in Figure~\ref{fig:cdrop}.

\begin{figure}[!ht]
	\centering
 	\includegraphics[width=0.8\textwidth]{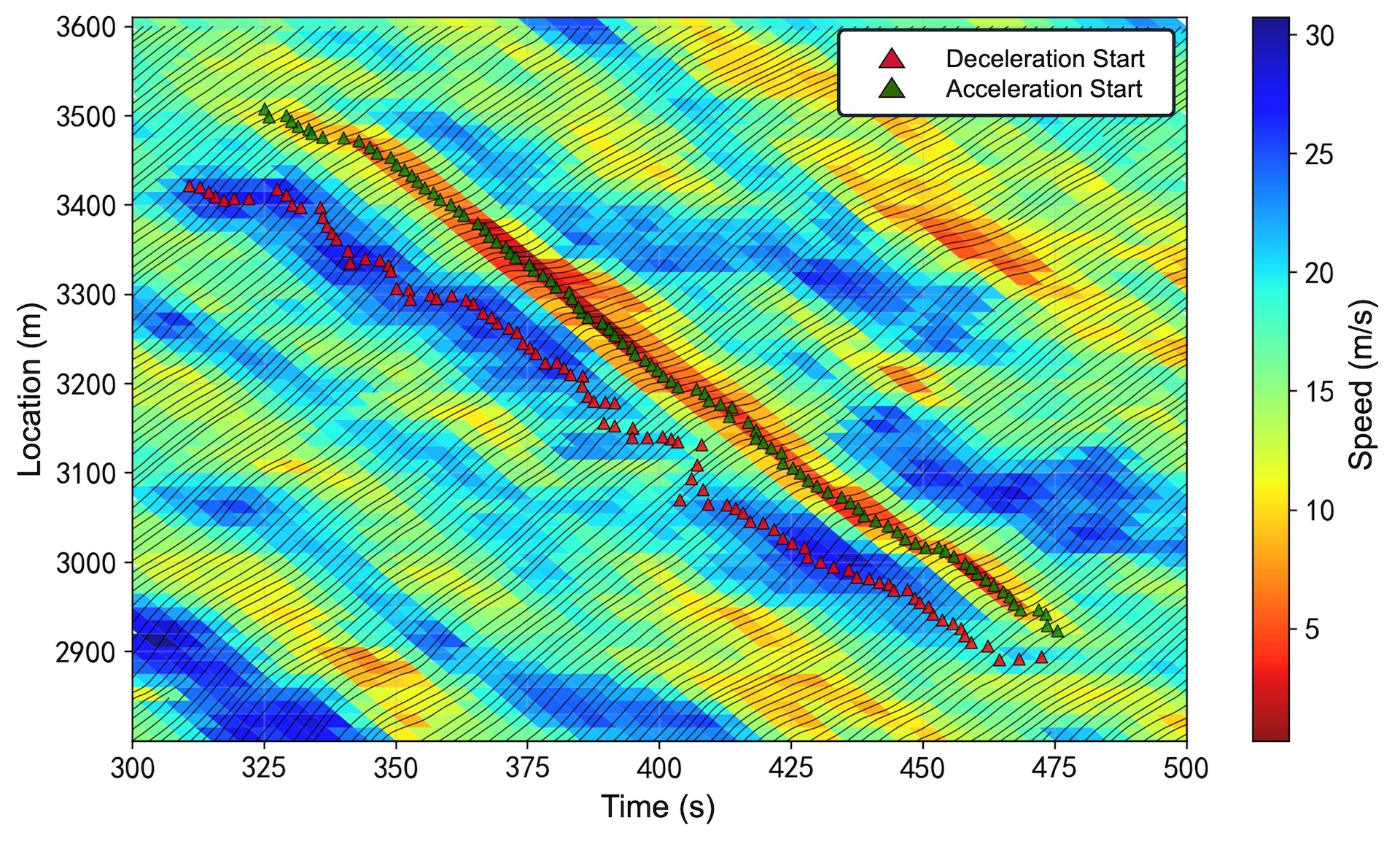}
	\caption{An example of a traffic oscillation covering its entire evolution, including formation, growth, and decay \citep{he2019constructing}.}
	\label{fig:oscillation}
\end{figure}

Traffic oscillations, characterized by alternating phases of acceleration and deceleration among vehicles, are another important traffic phenomenon that has been extensively studied using trajectory data. Previous studies have investigated the formation \citep{zheng2011freeway, chen2012microscopic} and growth patterns \citep{tian2016empirical} of traffic oscillations based on the NGSIM dataset. However, achieving a more complete understanding of traffic oscillations, particularly their decay processes, requires trajectory data with sufficiently large spatial and temporal coverage. Our dataset contains numerous oscillation cases that capture the entire evolution process, including formation, growth, and decay. An example illustrating the analysis of a complete oscillation process is shown in Figure~\ref{fig:oscillation}. In this figure, the onsets of deceleration and acceleration are detected using the wavelet transform method \citep{zheng2011freeway}. This dataset therefore provides a valuable basis for more comprehensive investigations of traffic oscillations, as well as for the development and evaluation of vehicle control strategies aimed at suppressing oscillatory traffic behavior.

\subsection{Traffic state estimation and trajectory reconstruction}

Trajectory data also provide an important basis for traffic state estimation and trajectory reconstruction. In practice, however, only a small fraction of trajectories may be observable due to limited probe vehicle penetration or sensing coverage. Estimating traffic states from sparse observations is therefore an important problem in traffic flow research. Datasets with long spatial coverage, extended observation duration, and continuous trajectories are particularly valuable for evaluating such methods.

\begin{figure*}[!ht]
    \centering
    \subfigure[]{\includegraphics[width=0.45\textwidth]{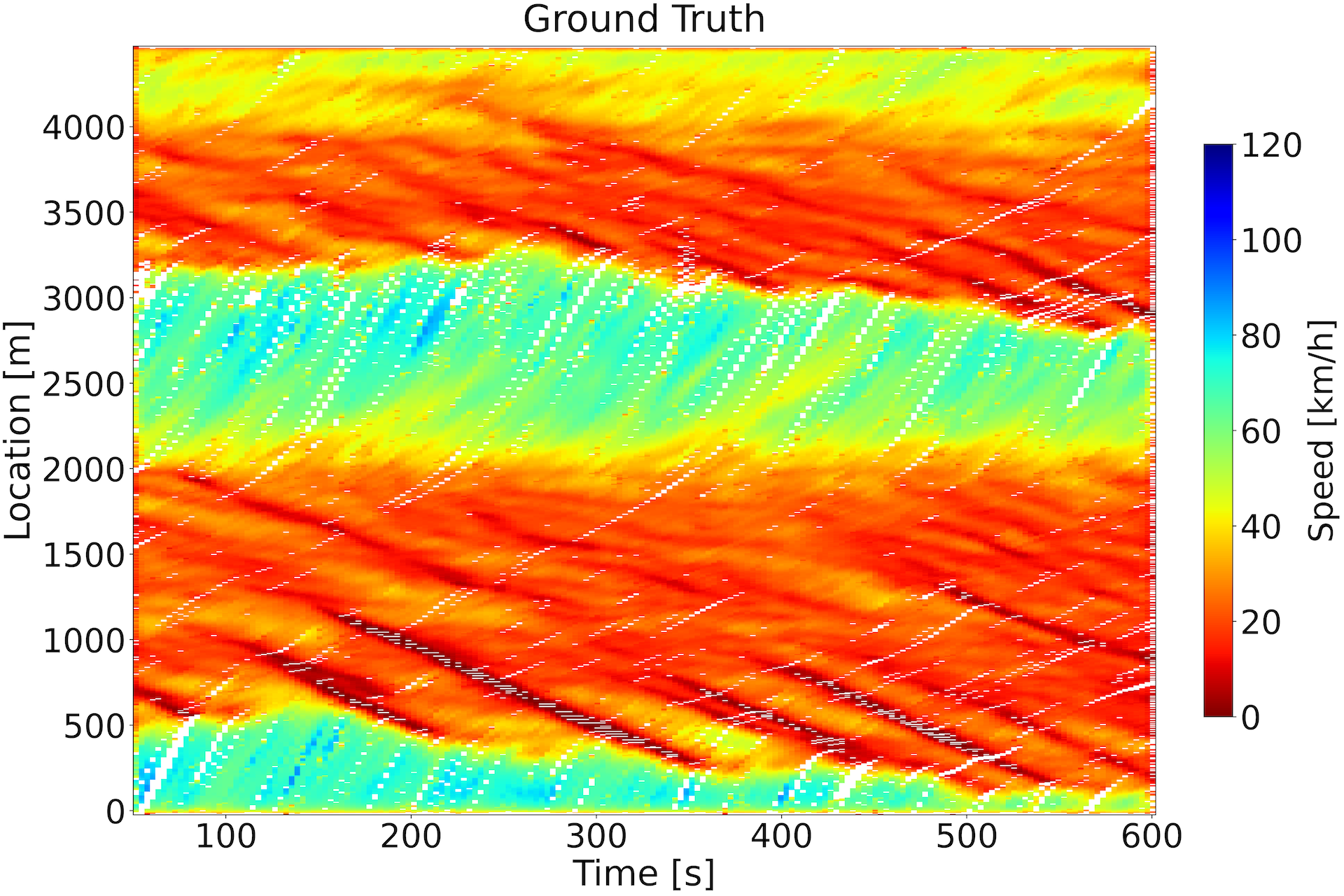}}
    \subfigure[]{\includegraphics[width=0.45\textwidth]{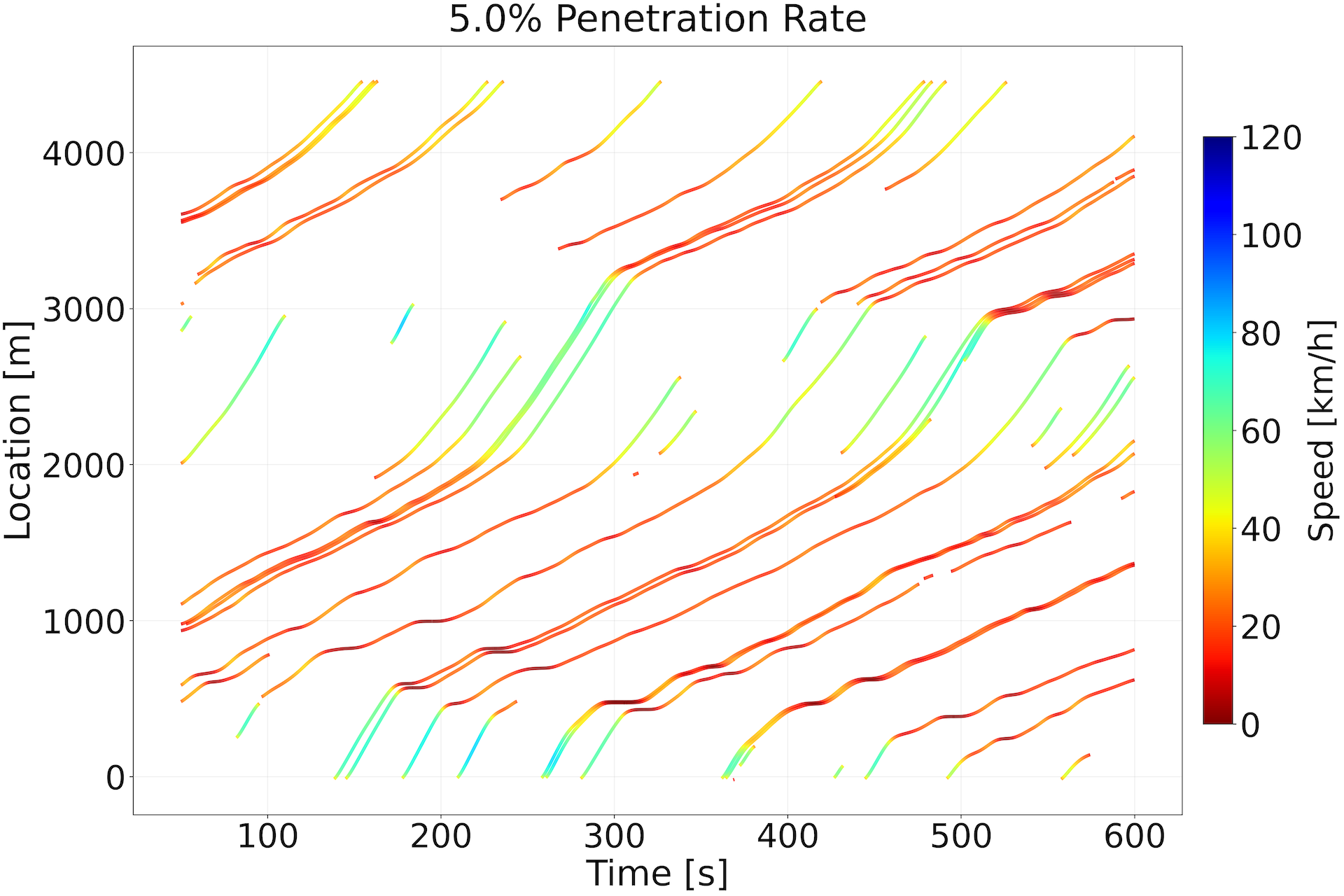}} \\
    
    \subfigure[]{\includegraphics[width=0.45\textwidth]{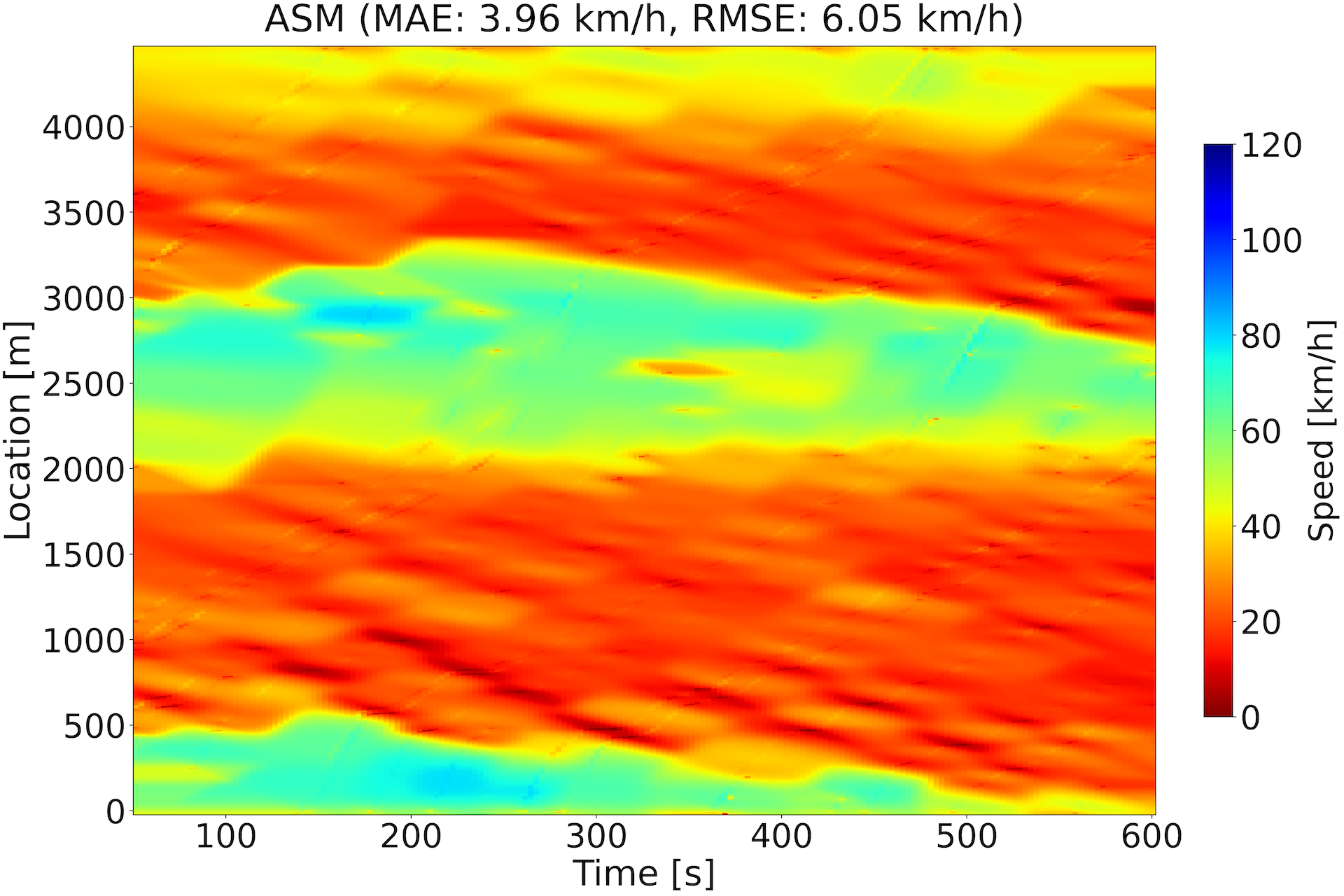}}
    \subfigure[]{\includegraphics[width=0.45\textwidth]{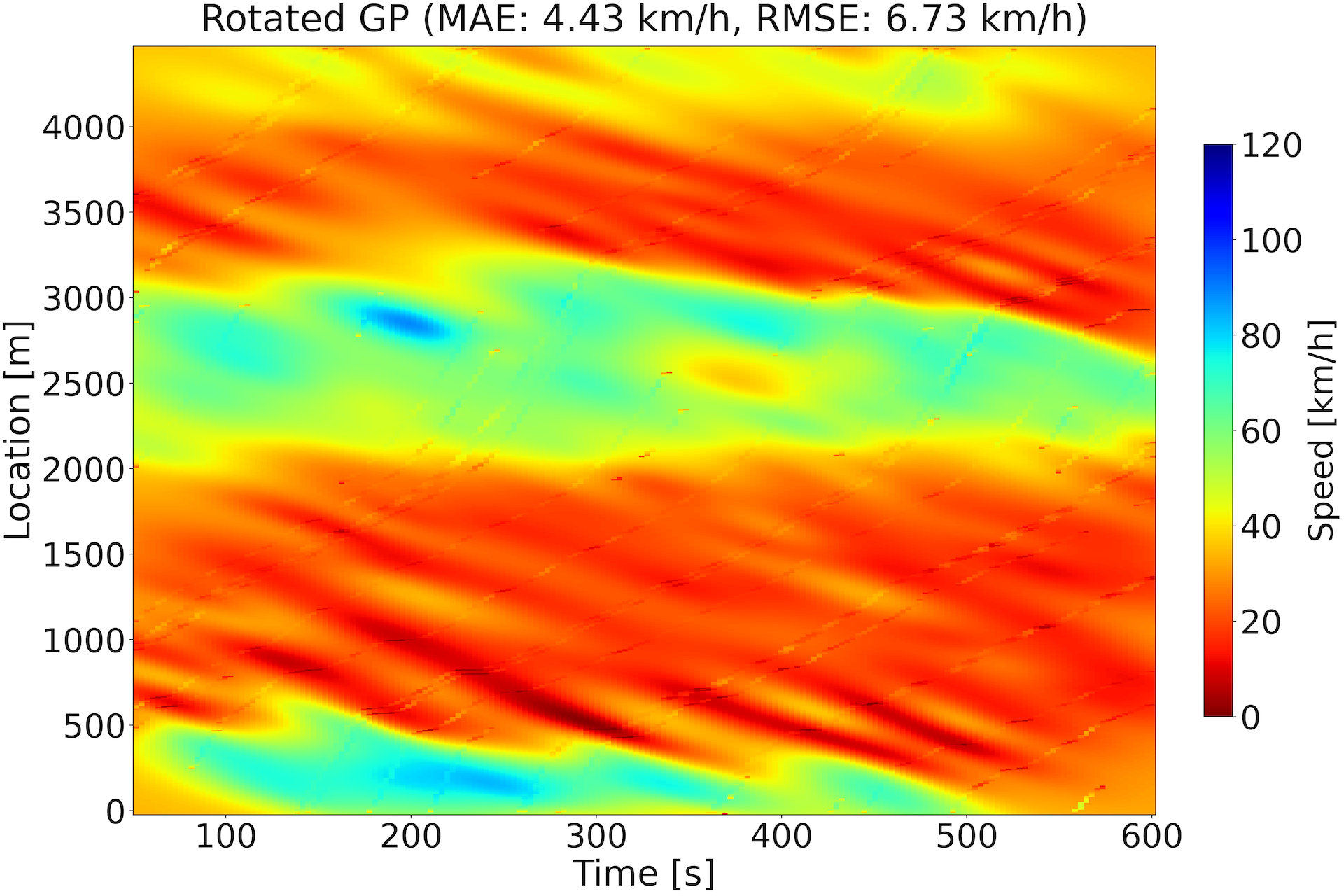}}
    
    \caption{(a) Ground-truth speed field. (b) 5\% observed trajectories. (c) Speed field reconstructed by ASM. (d) Speed field reconstructed by directional GP.}
    \label{fig:speed_comparison}
\end{figure*}

Fig.~\ref{fig:speed_comparison} shows an example with a 5\% trajectory penetration rate, i.e., only 5\% of vehicle trajectories are observed. Fig.~\ref{fig:speed_comparison}(a) presents the ground-truth speed field derived from complete trajectories based on Edie’s definition, and Fig.~\ref{fig:speed_comparison}(b) shows the sparsely observed trajectories. Fig.~\ref{fig:speed_comparison}(c) and Fig.~\ref{fig:speed_comparison}(d) present the speed fields reconstructed using the adaptive smoothing method (ASM) \citep{treiber2011reconstructing} and the rotated Gaussian process (GP) method \citep{wu2024traffic}, respectively. Both methods capture the main spatiotemporal traffic patterns, including the congested regions and the propagation of traffic waves. In this example, ASM achieves higher accuracy, with an MAE of 3.96 km/h and an RMSE of 6.05 km/h, compared with 4.43 km/h and 6.73 km/h for the rotated GP method.

To further evaluate the quality of the reconstructed macroscopic speed fields, vehicle trajectories were reconstructed using the integrated method \citep{chen2022integrated}. When the ASM-based speed field is used as input, the trajectory reconstruction achieves an MAE of 13.49 m and an RMSE of 18.96 m. In comparison, using the rotated GP-based speed field results in a higher MAE of 15.35 m and an RMSE of 21.41 m. These results indicate that, in this example, ASM provides better support for downstream trajectory reconstruction than the rotated GP method. This example further demonstrates that the proposed dataset can serve as a valuable benchmark for evaluating traffic state estimation and trajectory reconstruction methods over long spatial ranges under sparse-observation conditions.

\subsection{Modeling of vehicle behavior}

\begin{figure}[!ht]
	\centering
 	\includegraphics[width=0.96\textwidth]{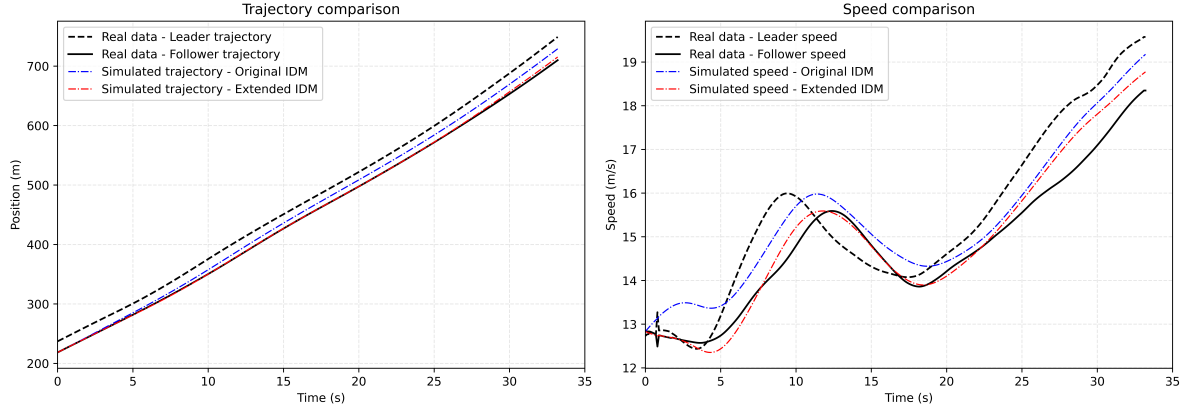}
	\caption{Comparison between the extended IDM and the original IDM in a car-following process.}
	\label{fig:extendedIDM}
\end{figure}

Vehicle behavior modeling relies heavily on empirical observations, and it is desirable for models to reproduce traffic phenomena observed in the field. Trajectory data therefore provide a valuable foundation for developing traffic flow models that more accurately capture traffic evolution. For example, in \citet{han2025capacity}, we found that the acceleration delays of hesitant vehicles, typically triggered by sharp decelerations or aggressive lane changes, are a major contributor to capacity drop. Such delayed acceleration behavior is not represented in classical car-following models. To address this limitation, an extended Intelligent Driver Model (IDM) was developed to explicitly capture the acceleration delay of hesitant vehicles. The extended model is able to reproduce this behavior effectively. An example comparing the extended IDM with the original IDM is shown in Figure~\ref{fig:extendedIDM}. In this example, both the extended IDM and the original IDM replicate the trajectory of the following vehicle, using calibrated parameters, based on the real trajectory of a leading vehicle. The comparison between the replicated trajectories and the real trajectory of the following vehicle shows that the extended IDM outperforms the original IDM in reproducing the behavior of such hesitant vehicles.

Building on behavior modeling at individual bottlenecks, long-distance trajectory data also enable the investigation of how driver behavior changes when vehicles travel along a corridor involving multiple bottlenecks. When traversing such corridors, drivers may maintain heightened attention and relatively small headway when approaching or passing bottlenecks, and exhibit more relaxed behavior after passing them. Such behavioral variations during corridor-level driving, particularly when multiple bottlenecks are involved, have rarely been examined, largely due to the lack of long-distance continuous trajectory data. This dataset contains a large number of continuous trajectories spanning more than 4.5 km, together with detailed information on surrounding vehicles. These data provide a basis for analyzing driver behavior variations over long-distance travel and for improving traffic management strategies at the corridor level.

While conventional car-following and lane-changing models were mostly developed based on physical principles, data-driven traffic flow models have attracted increasing attention in recent years. Such microscopic data-driven models typically require large amounts of trajectory data for training and validation. To date, the NGSIM dataset remains the most widely used source for this purpose \citep{parashar2025reassessing}. As a result, the validity of many data-driven models has been confined to a limited range of traffic scenarios represented in NGSIM. To examine the generalizability of these models when transferred to different traffic environments, additional trajectory data covering more diverse scenarios are required. Our dataset serves as a valuable supplement to existing open-source trajectory datasets in this regard, as it covers a variety of traffic scenarios, including merge, diverge, and weaving sections where mandatory lane changes occur, under different levels of congestion. In \citet{han2024modeling}, lane-change decision behavior and lane-change implementation processes were modeled using a parallel learning framework that integrates conventional behavioral models with deep learning components, based on the trajectory data, as illustrated in Figure~\ref{fig:lc}.

\begin{figure}[!ht]
	\label{fig:lc}
	\centering
	\includegraphics[width=0.85\textwidth]{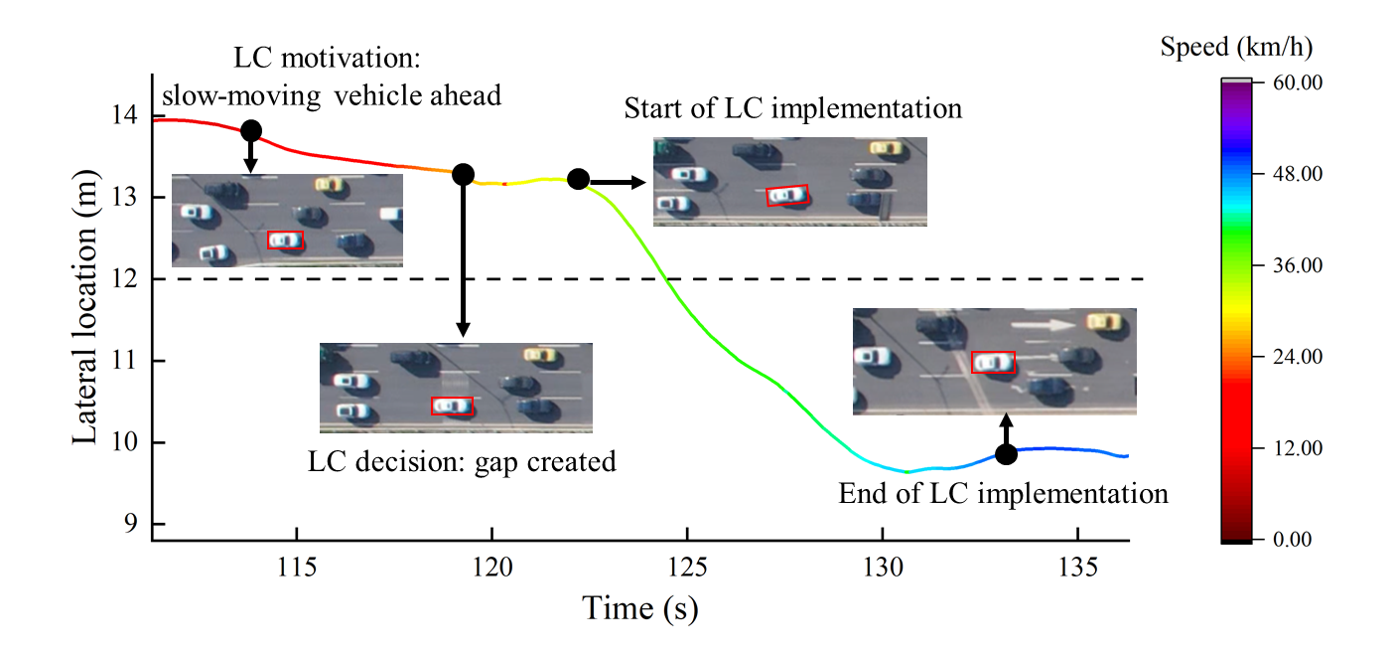}
	\caption{An example illustrating a vehicle’s lane-change process identified from the trajectory data.}\label{fig:lc}
\end{figure} 

\begin{figure*}[!h]
	\centering
	\subfigure[]{\includegraphics[width=0.6\textwidth] {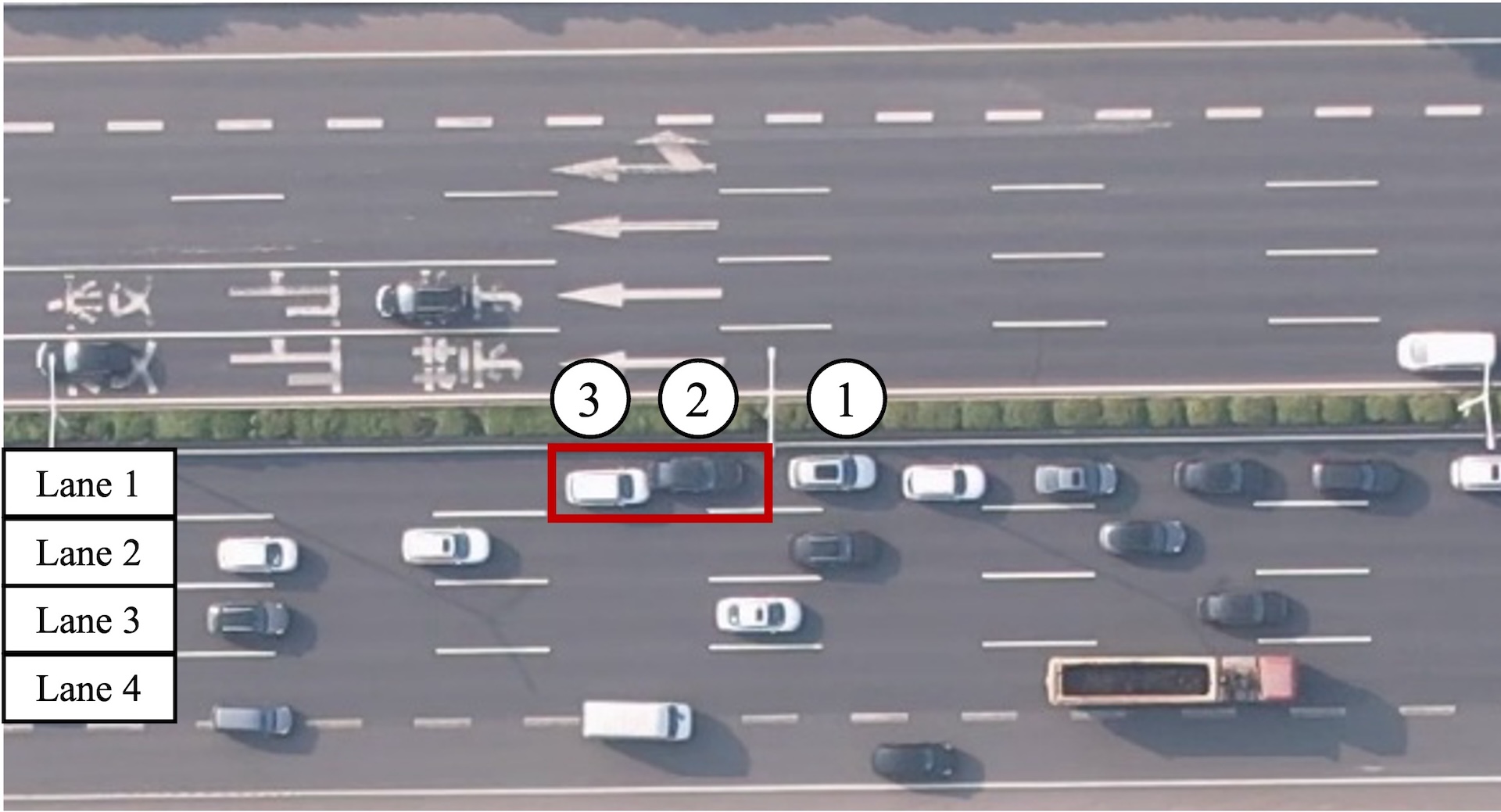}}
	\subfigure[]{\includegraphics[width=0.45\textwidth] {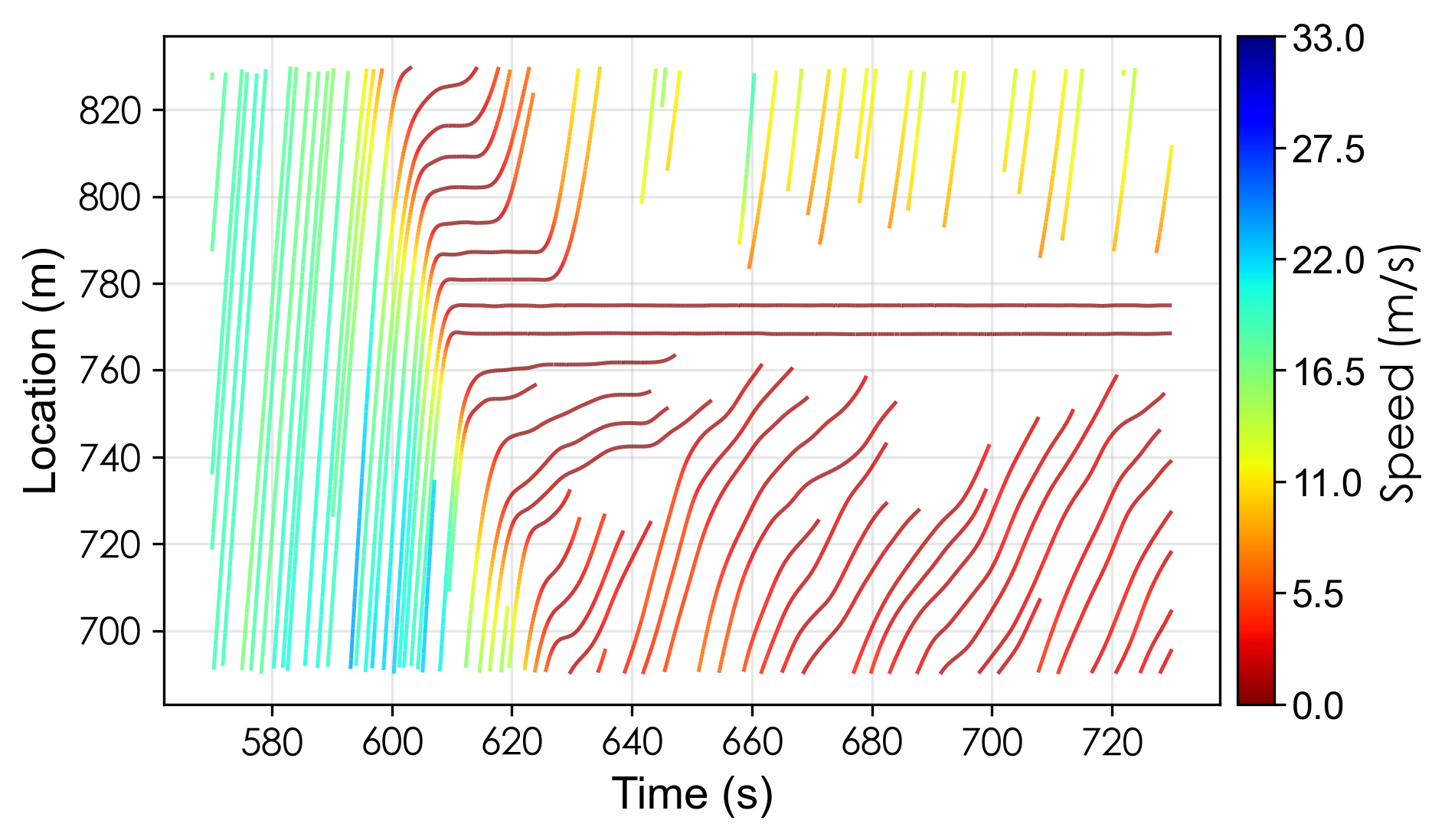}} 
    \subfigure[]{\includegraphics[width=0.45\textwidth] {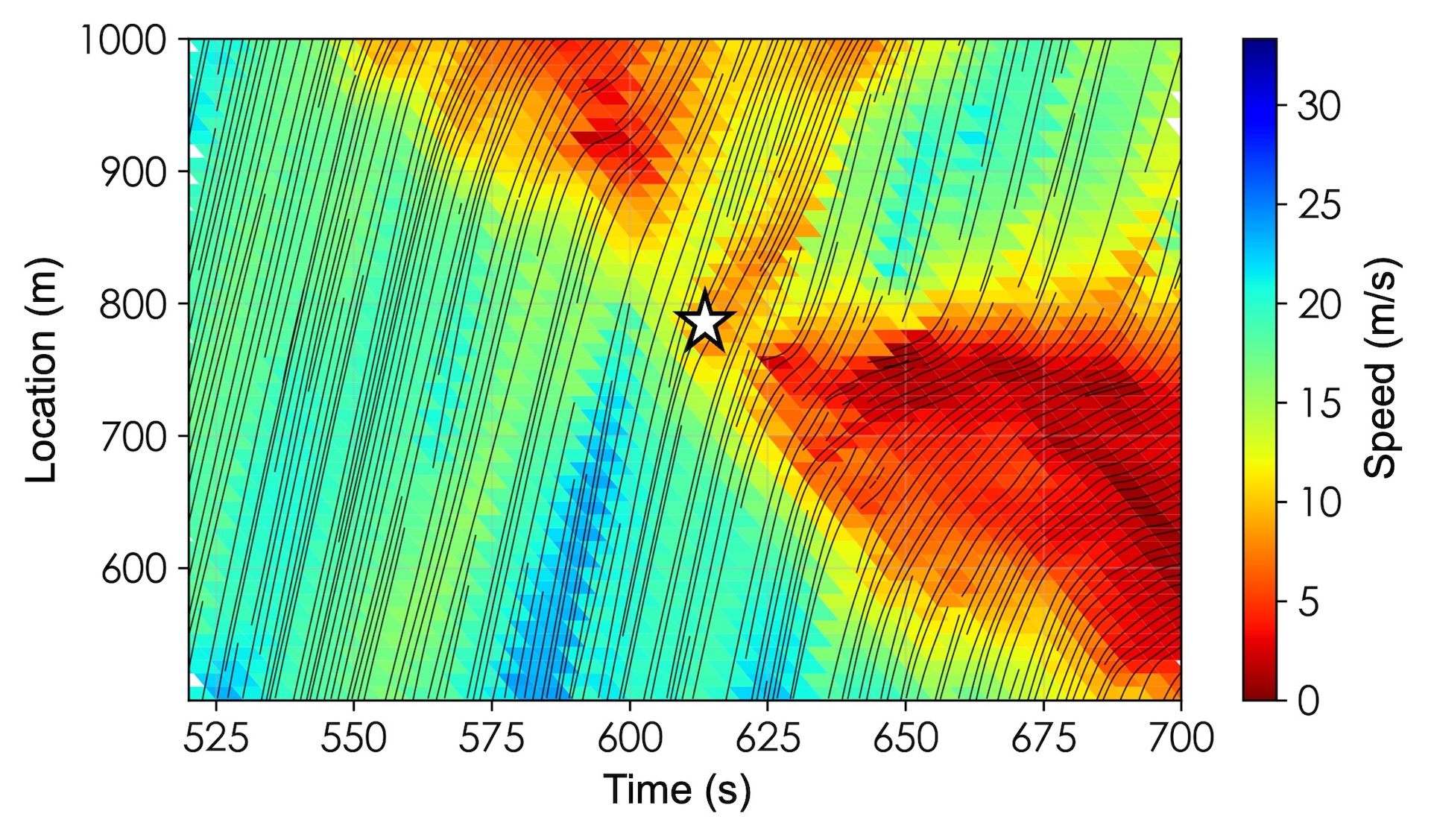} \label{fig:accident_speed}}
	\caption{ (a) Aerial view of the accident scene (b) Trajectory plot of lane 1 (c)  Speed plot of lane 2.}\label{fig:accident}
\end{figure*}

The dataset also contains several traffic accident events. Figure~\ref{fig:accident} (a) shows a UAV-captured aerial view at the moment of a rear-end collision. The red bounding box highlights the vehicle involved in the collision, and the vehicle ID is indicated by a circled label. This accident is identified as a rear-end collision in Lane 1, where the white vehicle (vehicle 3) failed to brake in time and collided with the black vehicle (vehicle 2) ahead. The corresponding spatiotemporal trajectories of vehicles in the accident lane are shown in Figure~\ref{fig:accident} (b), illustrating the evolution of traffic before and after the crash. Both vehicles involved in the collision came to a complete stop immediately after the impact. Subsequently, following vehicles began to change lanes in advance to avoid the obstructed section, resulting in a temporary local disturbance that propagated upstream along the traffic flow.

Figure~\ref{fig:accident} (c) presents the spatiotemporal speed heatmap of the adjacent lane (Lane 2). The star marker denotes the time and location of the accident. As illustrated, vehicles in the adjacent lane exhibit a rapid speed reduction shortly after the collision. This deceleration may be initially triggered by visual distraction, commonly referred to as rubbernecking, which describes speed reductions and increased headway variability caused by drivers’ attention to a nearby crash scene despite their own lane remaining unobstructed \citep{chung2013spatiotemporal}. The effect is further amplified by vehicles merging from the obstructed crash lane, leading to a more pronounced and sustained speed reduction in Lane~2. Such incident-induced behavioral changes can be further analyzed and modeled using this dataset.

\subsection{Network traffic analysis}

The presented dataset contains trajectories at the network level, covering both an urban network and its connected freeway. It can therefore support network-level traffic analysis, including congestion tracing \citep{yang2026traffic}, route choice analysis \citep{liu2023can}, and origin–destination estimation and prediction \citep{zhang2024physics}.

Network traffic flow modeling and control based on the macroscopic fundamental diagram (MFD) has been studied for about two decades since its introduction \citep{geroliminis2008existence}. However, most existing studies rely primarily on simulations, in which traffic demand, route choice, and driving behavior may deviate from real-world conditions. Empirical studies on MFDs are relatively limited and typically rely on aggregated data sources, such as loop detectors, which lack complete information on individual vehicle trajectories.

The pNEUMA dataset is the first open-source dataset providing complete trajectories of individual vehicles over an urban network, enabling detailed analysis of traffic state evolution from a network perspective. In contrast, our dataset includes trajectories from both an urban network and its connected freeway. Beyond characterizing MFDs for individual networks, the availability of synchronized trajectory data across these interconnected components enables the investigation of cross-network interactions. In practice, congestion on freeways may spill back into urban arterials, while urban signal control and route choice can significantly influence freeway demand. Such interactions are difficult to capture using traditional fixed-location sensors. The proposed dataset therefore provides a unique opportunity to study coordinated urban–freeway traffic dynamics and to evaluate integrated control strategies under realistic traffic conditions \citep{han2023coordinated}.

\begin{figure*}[!h]
	\centering
	\subfigure[]{\includegraphics[width=0.45\textwidth] {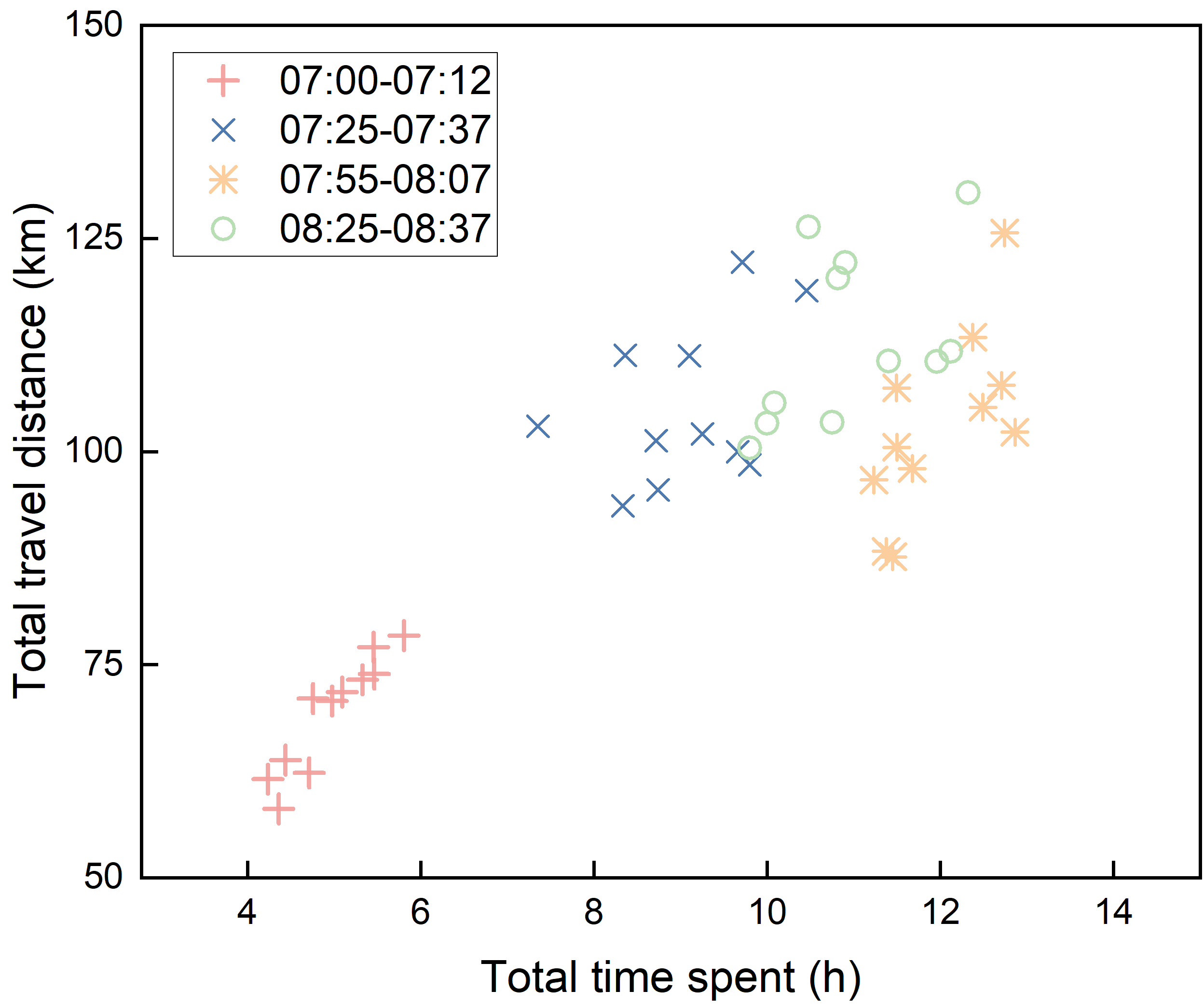}}
	\subfigure[]{\includegraphics[width=0.45\textwidth] {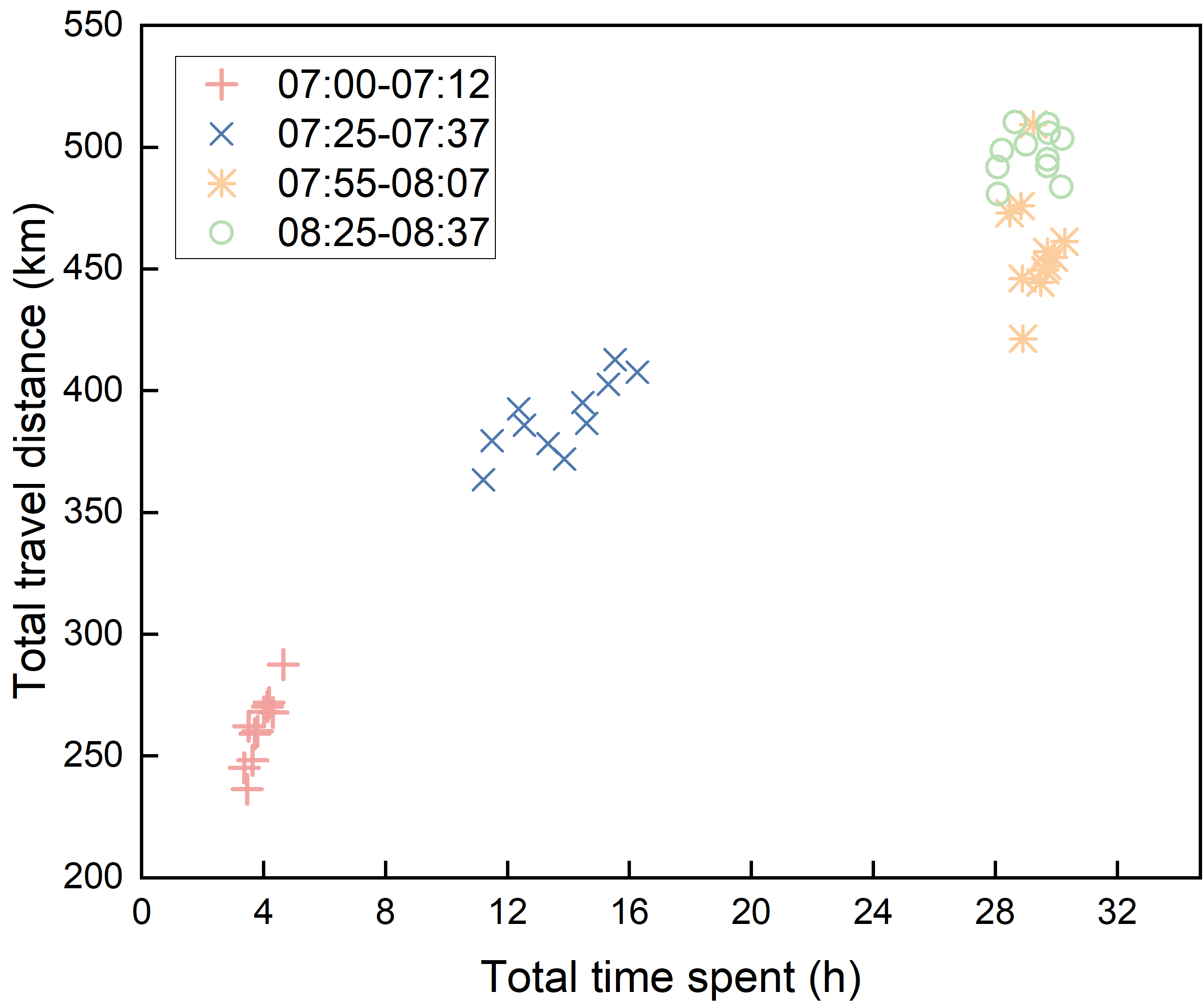}} 
	\caption{The MFDs of (a) the urban network, and (b) the freeway stretch.}\label{fig:mfd}
\end{figure*}

Figure~\ref{fig:mfd} illustrates the MFDs of the urban network and the freeway stretch across four time intervals on June~16,~2022, each lasting 12 minutes. Data were aggregated at a one-minute resolution. The first two intervals (07{:}00–07{:}12 and 07{:}25–07{:}37) correspond to traffic accumulation periods, during which total travel distance increases with total travel time. During the third interval (07{:}55–08{:}07), as traffic demand continues to increase, traffic efficiency degrades, as reflected by the reduction in total travel distance. The fourth interval represents the recovery phase.

\subsection{Intelligent vehicle-related research}

Research on intelligent vehicles using trajectory data primarily covers three aspects: trajectory prediction, decision-making, and scenario generation. Accurate trajectory prediction is essential for the safe deployment of connected and autonomous vehicles (CAVs) \citep{wei2024ki, geng2023dynamic, geng2023multimodal}. The inputs to trajectory prediction models typically include the historical and current states of the subject vehicle as well as those of surrounding vehicles. For CAV decision-making, a large body of recent research focuses on human-like driving behavior, where decision-making policies are trained using naturalistic driving data to enable CAVs to make decisions in a manner similar to human drivers \citep{zhou2026safedrive}. Such human-like, logic-driven behavioral decision-making and intelligent motion planning depend critically on accurately capturing how human drivers perceive, react to, and interact with surrounding traffic participants. Scenario generation is another important research direction, particularly for safety evaluation of autonomous driving systems under extreme or rare conditions \citep{ding2021multimodal}. The inputs to scenario generation models also rely heavily on real-world trajectory data.

While the NuScenes dataset \citep{caesar2020nuscenes}, which is collected using onboard perception sensors of autonomous vehicles, remains one of the most widely used datasets for trajectory prediction in autonomous driving tasks, drone-collected trajectory data offer several unique advantages. In particular, the interaction information captured in drone-based trajectory data is more comprehensive. It includes not only vehicles in the immediate vicinity of the subject vehicle, but also other surrounding vehicles that may influence its decision-making process. Moreover, compared with open-source datasets collected from vehicle-based perceptions, drone trajectory data contain a richer set of driving scenarios, especially under congested traffic conditions. For example, our dataset includes driving sessions during traffic oscillations, under accident conditions, as well as at non-typical intersections involving work zones. As such, it can serve as a useful supplement to existing datasets for CAV-related trajectory prediction and decision-making studies.

Nevertheless, to facilitate such applications, further data processing is required to reformulate the raw trajectories into standard session-based dataset formats, for example by converting Eulerian representations into Lagrangian coordinate systems, which remains an ongoing effort.

\section{Discussions}
\label{sec:dis}

This paper presents an open-source vehicle trajectory dataset collected from a UAV swarm experiment conducted in 2022 in Nanjing, China. While open science has attracted increasing attention in the transportation community and numerous open-source trajectory datasets have been released, the proposed dataset offers several distinctive features. First, it provides long-distance continuous trajectories of up to 4.5 km on a freeway, enabling in-depth investigation of traffic phenomena and their spatial and temporal evolution. Second, the data collection site covers an integrated network consisting of a long freeway corridor and parts of its connected urban network, facilitating traffic analysis and modeling from a network perspective.

Obtaining long-distance continuous trajectories is challenging, as trajectories from consecutive images must be accurately connected and occlusions caused by overpass bridges, which are common on freeways, must be reconstructed. These challenges are addressed in the presented dataset through the development of new techniques for trajectory reconstruction and connection. To the best of our knowledge, the freeway trajectories in this dataset represent the longest continuous trajectories, in terms of distance, among existing open-source trajectory datasets.

Beyond trajectory length, the integrated freeway–urban structure of the dataset enables analyses that are difficult to conduct using datasets collected from isolated sites. Continuous trajectories spanning multiple bottlenecks and roadway types allow researchers to examine how driving behavior, traffic states, and vehicle interactions evolve across different traffic environments within a single trip. This capability is particularly relevant for corridor-level traffic analysis, network-level modeling, and the study of behavioral adaptation under varying traffic conditions. Finally, this paper discusses a wide range of potential applications of the dataset beyond conventional traffic flow analysis and modeling. In the context of connected and autonomous driving, the dataset provides supplementary real-world scenarios for trajectory planning, decision-making, and scenario generation.

At the same time, it is important to clarify the scope of the dataset. The data were collected during morning peak hours on selected weekdays under favorable weather conditions and therefore primarily reflect recurrent congestion patterns. The dataset is not intended to represent all traffic conditions, such as nighttime driving or adverse weather scenarios. In addition, while the UAV-based observations provide detailed kinematic information, they do not include driver intent, vehicle automation status, or in-vehicle sensing data. These limitations should be considered when applying the dataset to specific research problems. The SWIFTraj dataset is not designed to replace existing open-source trajectory datasets, but rather to complement them. It will continue to be expanded and refined, and future releases may include additional sites, traffic conditions, and derived data products. The dataset is publicly available at the SWIFTraj website (\href{https://www.swiftraj.com}{https://www.swiftraj.com}).

\section*{Acknowledgment}

This research is supported by the National Natural Science Foundation of China (No.52525204, No.52232012).

\appendix
\section{Data Format}
\label{sec:data_format}

The dataset is organized into two main components: metadata and trajectory data. The metadata component provides file-level and site-level information required for interpreting and using the trajectory data. The trajectory data component contains detailed spatiotemporal records of individual vehicles observed in the scene. Table~\ref{tab:dataset_metadata} lists the metadata fields, while Table~\ref{tab:trajectory_fields} details the fields associated with vehicle trajectories.


\begin{table}[htbp]
\centering
\caption{Definitions of metadata fields.}
\label{tab:dataset_metadata}
\begin{tabular}{p{5cm} p{2.0cm} p{8.0cm}}
\hline
\textbf{Field} & \textbf{Type} & \textbf{Description} \\
\hline
data\_file\_name
& string
& Name of the data file (e.g., ``Hurong\_20220617\_A1\_F1'') \\

location\_id
& string
& Identifier of the data collection site (e.g., ``A1'') \\

location\_name
& string
& Textual description of the camera capture location, recommended in the order of location--city--province/state--country (e.g., ``Hurong--Nanjing--Jiangsu--China'') \\

frame\_interval
& float
& Time interval between consecutive frames, in seconds (e.g., 0.1) \\

start\_timestamp\_ms
& int64
& Unix timestamp at the start of data collection, expressed in milliseconds (e.g., 1655420390457) \\

start\_datetime
& string
& Human-readable start time of data collection (format: YYYY-MM-DD HH:MM:SS; e.g., ``2022-06-17 06:59:50'') \\

total\_duration
& float
& Total duration of the trajectory recording, in seconds (e.g., 300.0) \\

timestamp\_timezone
& string
& Time zone used to interpret the Unix timestamps (e.g., ``Asia/Shanghai'') \\

spatial\_unit
& string
& Unit of metric spatial coordinates (e.g., ``m'') \\

dataset\_version
& string
& Version identifier of the dataset release (e.g., ``1.0.0'') \\

lane\_sequence\_to\_movement\_map
& dict
& Mapping between each movement ID and its movement direction. (intersection data only), (e.g., \{``1-3-20'': ``Right-turn''\}) \\

total\_vehicle\_count 
& int
& Total number of vehicle trajectories contained in the data file (e.g., 12345) \\

unique\_lane\_ids
& list[int]
& Set of all lane identifiers appearing in the data file (e.g., [1, 2, 3, 20]) \\
\hline
\end{tabular}
\end{table}

\begin{table}[htbp]
\centering
\caption{Definitions of trajectory fields.}
\label{tab:trajectory_fields}
\begin{tabular}{p{2.5cm} p{2.5cm} p{10.0cm}}
\hline
\textbf{Field} & \textbf{Type} & \textbf{Description} \\
\hline
vehicle\_id 
& int 
& Unique identifier of a vehicle trajectory \\

vehicle\_class 
& string 
& Vehicle category (e.g., Car or Truck) \\

vehicle\_width 
& float 
& Vehicle width \\

vehicle\_length 
& float 
& Vehicle length \\

frame\_index 
& list[int] 
& Sequence of frame indices associated with the trajectory \\

frenet\_s 
& list[float] 
& Longitudinal position of the vehicle center in the Frenet coordinate system (expressway data only) \\

frenet\_d 
& list[float] 
& Lateral offset of the vehicle center in the Frenet coordinate system (expressway data only) \\

frenet\_s\_speed 
& list[float] 
& Longitudinal speed in the Frenet coordinate system (expressway data only)\\

frenet\_d\_speed 
& list[float] 
& Lateral speed in the Frenet coordinate system (expressway data only) \\

frenet\_s\_accel 
& list[float] 
& Longitudinal acceleration in the Frenet coordinate system (expressway data only) \\

frenet\_d\_accel 
& list[float] 
& Lateral acceleration in the Frenet coordinate system (expressway data only) \\

lane\_id 
& list[int] 
& Identifier for the lane (freeway) or the corresponding road segment or intersection (urban road) in which the vehicle center is located at each frame.\\

pixel\_x 
& list[float] 
& Per-frame horizontal coordinate of the vehicle center in pixel space \\

pixel\_y 
& list[float] 
& Per-frame vertical coordinate of the vehicle center in pixel space \\

ground\_x 
& list[float] 
& Per-frame horizontal coordinate of the vehicle center in the local ground coordinate system \\

ground\_y 
& list[float] 
& Per-frame vertical coordinate of the vehicle center in the local ground coordinate system \\

pixel\_corners 
& list[list[float]] 
& Per-frame pixel coordinates of the four vehicle corners, ordered clockwise \\

ground\_corners 
& list[list[float]] 
& Per-frame ground coordinates of the four vehicle corners, ordered consistently with pixel\_corners \\

is\_imputed 
& list[int] 
& Per-frame indicator of whether the state is observed (0) or imputed/reconstructed (1) \\
\hline
\end{tabular}
\end{table}

\clearpage
\bibliographystyle{elsarticle-harv} 
\bibliography{main}

\end{document}